\documentclass[12pt]{article}
\usepackage[sc]{mathpazo}
\usepackage{xy}
\usepackage{url}
\usepackage{authblk}
\usepackage{hyperref}
\usepackage{graphicx}
\usepackage{amssymb,amsmath,amsthm}
\usepackage{bbold}
\usepackage{bm}
\def\RPt{{\bR {\mathrm P}^2}}
\def\RPf{{\bR {\mathrm P}^4}}
\def\QPt{{\bQ {\mathrm P}^2}}
\def\QPT{{\bQ {\mathrm P}^3}}
\def\emb{\hookrightarrow}

\def\mucube{\hbox{$\mu$-cube}}
\def\mupar{\hbox{$\mu$-parallelepiped}}
\def\mupars{\hbox{$\mu$-parallelepipeds}}
\def\c{c}

\def\irr{\mathop{irr}}

\def\sp{\mathop{span}}

\def\rk{\mathop{rk}}
\def\zone{\cD}
\def\fc{{\frak c}}
\def\fd{{\frak d}}
\def\cE{{\cal E}}
\def\rational{{\bQ}}
\def\B{{\bf B}}
\def\p{{\bf p}}
\def\l{{\bf l}}

\def\bH{{\boldsymbol{B}}}
\def\bT{{\mathbb T}}
\def\boC{{\boldsymbol{C}}}

\def\bZ{\mathbb{Z}}
\def\bC{\mathbb{C}}
\def\bQ{\mathbb{Q}}
\def\bR{\mathbb{R}}
\def\bS{\mathbb{S}}

\def\Rt{{\mathbb{R}^2}}

\def\cH{{\cal H}}
\def\cM{M}

\def\low#1{l_{#1}}
\def\upp#1{u_{#1}}

\def\cC{{\cal C}}
\def\cD{{\cal D}}
\def\cE{{\cal E}}

\def\cK{{\cal K}}
\def\cH{{\cal H}}
\def\cL{{\cal L}}
\def\cM{{\cal M}}

\def\cF{{\cal F}}
\def\cU{{\cal U}}

\def\rf{{f_p}}

\def\real{\bR}

\newtheorem{definition}{Definition}
\newtheorem{theorem}{Theorem}

\newtheorem{example}{Example}

\newtheorem{remark}{Remark}

\newtheorem{proposition}{Proposition}
\newtheorem{conjecture}{Conjecture} 

\usepackage{array}
\newcolumntype{S}{>{\centering\arraybackslash} m{.475\linewidth}}
\newcolumntype{T}{>{\centering\arraybackslash} m{10.5cm}}
\newcolumntype{U}{>{\centering\arraybackslash} m{1.5cm}}
\title{%
  Quasiperiodic functions on the plane\\ and electron transport phenomena.
}
\author[1]{Roberto De Leo}
\affil[1]{Department of Mathematics,\par
  Howard University,\par
  Washington, DC 20059
}
\author[2]{Andrei Ya. Maltsev}
\affil[2]{Landau Institute for Theoretical Physics,\par
  Russian Academy of Sciences,\par
  142432 Chernogolovka, Moscow region
}
\begin{document}
\maketitle
\begin{abstract}
  While quasiperiodic functions in one variable appeared in applications since Eighteen hundreds,
  for example in connection with the trajectories of mechanical systems with $2n$ degress of freedom
  having $n$ commuting first integrals, the first applications of multivariable quasiperiodic functions
  were found only in Seventies, in connection with solitonic solutions of the KdV equation.
  Later, several other physical applications were found, especially in connection with Solid State
  Physics, in particular with the electron transport phenomena. In this article we reformulate,
  specifically in terms of the topology of level sets of quasiperiodic functions on the plane, some
  fundamental theoretical results found in Eighties and Nineties, then we review the physical
  models of electron transport and their connections with quasiperiodic functions and finally
  we present some old and new numerical results on the level sets of some specific family of
  quasiperiodic functions, some of which related to the magnetoresistance in normal metals.
\end{abstract}
\section{Introduction}
%
The canonical projection $\pi_{k}:\bR^k\to\bT^k\simeq\bR^k/\bZ^k$ determines 
a canonical linear embedding $C^\infty(\bT^k)\hookrightarrow C^\infty(\bR^k)$ defined by
$F\mapsto \pi_k^*F=F\circ\pi_k$ whose image is the set of {\sl periodic} functions on $\bR^k$.
{Quasiperiodic (QP) functions (sometimes also called {\em conditionally periodic} functions in literature)
  are produced by the following small perturbation of the procedure above:}
\begin{definition}
  Let $A_{n,k}$ be the set of all affine embeddings $\psi:\bR^k\hookrightarrow \bR^n$, $k<n$.
  Then the image $QP_{k,n}$ of the map $A_{n,k}\times C^\infty(\bT^n)\to C^\infty(\bR^k)$ defined by
  $$(\psi,F)\mapsto F_\psi=(\pi_{n}\circ\psi)^*F$$ is the set of {\sl quasiperiodic} functions
  on $\bR^k$ with {\sl at most} $n$ quasiperiods. The number of quasiperiods of $f$ is
  the smallest $n$ for which $f\in QP_{k,n}$ or, equivalently, the rank of the sublattice of
  $(\bR^k)^*$ generated by $\{\epsilon_i=\psi^*\eta_i\}$, $i=1,\dots,n$, where $\eta_i$ is any base for $(\bR^n)^*$.
\end{definition}
%
\begin{example}
  The function in one variable $f(x)=\cos(2\pi x)+\cos(\sqrt{2}\,2\pi x)$ is not periodic but it can be written as
  the restriction of the periodic function in two variables $\varphi(x,y)=\cos(2\pi x)+\cos(2\pi y)$ to the straight
  line $y=\sqrt{2}x$. Since $\sqrt{2}$ is not rational, $f$ is not periodic and so is a QP function
  in one variable with two quasiperiods. Its group of frequencies is generated by the covectors
  $\epsilon_1:x\mapsto x=\eta_1(x,\sqrt{2}x)$ and $\epsilon_2:x\mapsto\sqrt{2}x=\eta_2(x,\sqrt{2}x)$.
\end{example}
%
%
By abuse of notation, we will sometimes denote by $\psi$ the $k$-plane $\psi(\bR^k)$.
Note that, for any $\psi\in A_{n,k}$, the closure of $\pi_n(\psi(\bR^k))$ in $\bT^n$ is
an embedded torus of some dimension between $k$ and $n$. We say
that $\psi$ is {\sl rational} when $\overline{\pi_n(\psi(\bR^k))}=\bT^k$ and {\sl fully irrational}
when $\overline{\pi_n(\psi(\ell))}=\bT^n$ for every straight line $\ell\subset\bR^k$.

When $\psi_1,\psi_2\in A_{n,k}$ differ just by a constant, namely $\psi_1(x)=\psi_2(x)+y_0$
for some $y_0\in\bR^n$ and all $x\in\bR^k$,  
their images are parallel and we say that they are {\sl siblings}.
In this case, given any $F\in C^\infty(\bT^3)$ and provided that $\psi_1$ (and so $\psi_2$) is not rational,
$F_{\psi_1}$ and $F_{\psi_2}$ are {\sl almost} the translate of each other, namely 
for every $\varepsilon>0$ there is a $y_\varepsilon\in\bR^n$ s.t.
$$\sup_{y\in\bR^n}|F_{\psi_1}(y+y_\epsilon)-F_{\psi_2}(y)|<\varepsilon.$$

Quasiperiodic functions were introduced in the mathematical literature at the end of XIX century by the Latvian mathematician
P. Bohl~\cite{Bohl93}, in the context of the theory of differential equations, and by the French astronomer
E. Esclangon~\cite{Esc04}, who introduced the terminology. They become widely known to the mathematical
community, though, only in Nineteen-Twenties through the works of H. Bohr~\cite{Bohr26}, A. Besikovich~\cite{Bes26}
and S. Bochner~\cite{Boc27a,Boc27b} as a particular case of the more general theory of {\em Almost Periodic} functions:
a function is almost periodic when it can be written as a converging (in the supremum norm) series of trigonometric polynomials
and it is quasiperiodic when the set of all periods of the summands in the series is finite.

Applications of QP functions in one variable have been known even before they were formally defined since they
appear naturally in solutions of Ordinary Differential Equations and play an important role in Classical Mechanics:
as it was pointed out first by J. Liouville in a short note back in 1853~\cite{Lio55} and proved in full generality
later by V.I.~Arnold (e.g. see~\cite{AG90}), when a Hamiltonian system with $2n$ degrees of freedom admits $n$
first integrals in involution, it is possible to define angle-action variables so that, on any compact level set,
the solutions of the equations of motion are QP functions in one variable with at most $n$ quasiperiods.

The first application of multivariable QP functions appeared in literature only in Seventies and is due to
S.P. Novikov that, in his first and seminal work on solitons~\cite{Nov74}, showed that QP functions in more
than one variable appear naturally as solutions of {\em Completely Integrable PDEs}, in particular in case
of the KdV equation (see also~\cite{DN74,DNM76,DKN90} for more details and further developments).

The second field where multivariable QP functions play a major role is Solid State Physics.
A first occurrence, to which most of this article is dedicated, goes back to Fifties when I. Lifshitz, M.Ya. Azbel and
M.I. Kaganov, from the Kharkov-Moscow school of solid state physics,
in order to find a model able to describe several phenomena explained until then through some artificial assumptions,
developed a theory of conductivity in metals under a strong magnetic field based solely on the semiclassical
model~\cite{LAK56,LAK57,LAK73}.

Recall that, in this model, (quasi-)electrons are treated as classical particles
except for the (fundamental!) difference that their (quasi-)momenta belong to $\bT^3$ rather than $\bR^3$ and,
under a {\sl constant} magnetic field $\B=B^idp_i$, they obey the equation
$\dot p_i=\{p_i,\varepsilon(p)\}_\B\,,\; i=1,2,3$, where $\varepsilon(p)$ is the Fermi energy function
and $\{\}_\B$ is the ``magnetic Poisson bracket'' given by $\{p_i,p_j\}_\B = \epsilon_{ijk}B^k$. 
In~\cite{Nov82} Novikov pointed out that the function $b(p) = B^ip_i$ is a {\sl multivalued} Casimir for $\{\}_B$
and therefore the orbits of the solutions of the corresponding Hamiltonian equations are given, in the universal cover
$\bR^3$, by the level curves of a QP function in two variables with three quasiperiods. 


The theory of Lifshitz, Azbel and Kaganov predicted that the magnetoresistance would
depend qualitatively on the topology of the orbits of quasi-electrons' momenta that, therefore,
would be \textit{\textbf{observable}}.
Many experiments followed and fully confirmed the correctness of this model
(see the references in Sec.~\ref{sec:Mal} and Fig.~\ref{fig:nm})
but the theoretical efforts in this direction stopped, after about a decade, because no method
was found to predict the topology of the orbits for a general Fermi Surface, until the recent 
fundamental results by Novikov and his topological school (I. Dynnikov, A. Zorich and S. Tsarev)
in Eighties and Nineties (see Sec.~\ref{sec:QP}), which made possible several recent fundamental
theoretical advances by Novikov and the second author (see Sec.~\ref{sec:Mal}).

A second occurrence arose in Eighties
after the discovery
(that granted him the Nobel prize in 2011) by D. Schechtman~\cite{SBGC84} of states of matter with a symmetry
not corresponding to any crystal lattice (namely a discrete sublattice of $\bR^2$ or $\bR^3$) but rather to the
one of a quasicrystal, an object discovered in mathematics just a few years earlier by R. Penrose~\cite{Pen79}.
The relation betweeen quasicrystals and quasiperiodicity is that a quasicrystal can be seen as a quasiperiodic
tiling of the space. More precisely, a quasicrystal in $\bR^k$ is a collection of a countable number of closed polytopes
whose union is the whole space, whose pairwise intersection is either empty or an entire lower dimension subpolytope and such that:
{\em i.} modulo translations, there are only a finite number of them; {\em ii.} all functions that are constant in the interior
of the polytopes and assume the same value on those that can obtained by translations from one another are quasiperiodic.
Important contributions to the relation between quasicrystals and quasiperiodic functions were given by Novikov and
his school (Le Tu, Piunikhin, Sadov)~\cite{LPS93} and Arnold~\cite{Arn90}.
Analogously to the definition
of QP functions, all these tilings can be obtained as the intersection of a periodic tiling of $\bR^n$ with
a $k$-plane~\cite{Arn90}.

A third occurrence, again in Eighties, is the diffusion of particles in a magnetic field. {The interaction of a particle with a plane wave
  in a transverse magnetic field leads to an equation of the type
  $$\ddot x+\omega_\B^2 x=-E_0\sin(kx-\omega t)\,,$$
where $\omega_\B$ is the cyclotron frequency~\cite{ZZSUC86}.} When $\omega/\omega_\B$ is rational, the solutions of this equation
foliate the phase $(x,p_x)$ with countably many disjoint islands of closed orbits embedded in a sea of open ones,
allowing particles to diffuse arbitrarily far in the high energy region. Arnold showed~\cite{Arn90} that, in particular cases,
the solutions orbits can be approximated by the level sets of the QP Hamiltonian $H_\psi$ in two variables
with $n$ quasiperiods, where $H(\alpha^1,\dots,\alpha^n)=\sum_{l=1}^n\cos(2\pi\alpha^l)$ and $\psi\in A_{n,2}$ is 
fully irrational, explaining this way the similitude between quasicrystals and patterns
detected in the distribution of the islands of closed orbits. 

A fourth and last occurrence we want to mention is a theoretical prediction~\cite{Mal04,NM06} by the second author of this article based on a
semiclassical description of a 2-dimensional electron gas (2DEG) subject to a weak quasiperiodic potential.
A 2DEG is a semiconductor structure where the motion of electrons in one direction is somehow constrained
(and therefore quantized) so that in many phenomena only the projection of the momentum in the plane perpendicular to the
constrained direction plays a role and so the system can be considered 2-dimensional.
According to quasiclassical analysis (see e.g. \cite{Fer88,Bee89}), 
when a constant magnetic field $B$ and an electric field $E=dV$ are applied to a 2DEG, the drift
of the center of the cyclotron orbits $r$ satisfies, in appropriate units, the equations
$$\dot r_i = \frac{1}{\|\B\|^2}\{r_i, V^{eff}_\B\}_\B\,,$$
where $V^{eff}_\B$ is an effective potential depending on $V$ and $\B$.
We can always reduce to the case $\B=B_zdz$ and consider $V=V(x,y)$. {QP potentials in two variables with any number of
quasiperiods can be artificially generated and, in turn, the topological properties of the trajectories can be detected experimentally
by measurements of the magnetoresistance of the 2DEG.}
%
\section{QP functions on $\bR^2$ with 3 and 4 quasiperiods}
\label{sec:QP}
The study of the topology of the level sets of quasiperiodic functions in two variables with three quasiperiods
was posed in 1982 by S.P. Novikov in his celebrated paper extending Morse theory to multivalued functions~\cite{Nov82}
as a non-trivial application of his theory and intensively studied since then analytically and numerically
by Novikov~\cite{Nov91,Nov95,Nov99,Nov00} and his pupils A.V. Zorich~\cite{Zor84,Zor88}, S.P. Tsarev,
I.A. Dynnikov~\cite{Dyn92,Dyn96,Dyn97,Dyn99,Dyn08,DD09} and the first author~\cite{DeL00,DeL03a,DeL03b,DeL05a,DeL06,DD09}.
Lately important contributions were also given by A. Skripchenko~\cite{Skr13} jointly with Dynnikov~\cite{DS15} and
A. Avila and P. Hubert~\cite{AHS16,AHS16b}. The following theorem contains the most important topological results
relative to the case of 3 quasiperiods:
%
\begin{definition}
  We say that a planar open curve $\gamma$ is strongly asymptotic to a straight line $\ell$ when
  $\gamma$ lies inside a finite width strip parallel to the straight line and a generic line
  transversal to $\ell$ cuts $\gamma$ in an odd number of points. We say that $\gamma$ is a $B$-section
  if it lies on a plane perpendicular to a direction $B$.
\end{definition}
\begin{theorem}[Zorich~\cite{Zor84}, Dynnikov~\cite{Dyn99}]
  For every generic function $F\in C^\infty(\bT^3)$ there exist two continuous functions
  $\low{F},\upp{F}:\RPt\to\bR$, with $\low{F}\leq \upp{F}$ pointwise, 
  and a locally constant function $\ell_F:\cD_F=\{\low{F}<\upp{F}\}\subset\RPt\to\QPt$ such that:
  \begin{enumerate}
    \item $\QPt\subset\cD_F$;
    \item either $\cD_F=\RPt$, and then $\ell_F$ is constant, or $\ell_F$ assumes infinitely many values, and then
      $\cE_F=\RPt\setminus\cup_{l\in\ell(\cD_F)}\overline{\cD_{l,F}}$, where $\cD_{l,F}=\{\ell_F(B)=l\}$, is non-empty
      and uncountable;
    \item all planar sections of $M_c=\pi_3^{-1}\left(\{F=c\}\right)$ perpendicular to a fully irrational direction
      $B$ are closed if $c\not\in[\low{F}(B),\upp{F}(B)]$;
    \item if $B\in\cD_F$, then there are open non-singular $B$-sections on all level surfaces
      $M_c$ with $c\in\left[\low{F}(B),\upp{F}(B)\right]$ and, if $B$ is fully irrational, they are all strongly asymptotic
      to straight lines with direction $B\times\ell_F(B)$ ;
    \item if $B\in\cE_f$ and $e$ is the common value of $\low{F}(B)$ and $\upp{F}(B)$, then there are open $B$-sections
      of $M_e$ and none of them is strongly asymptotic to a straight line.
  \end{enumerate}
  \end{theorem}
As a corollary of this theorem, we get the following claim about the level sets of QP functions:
%
\begin{theorem}
  Let $q:\bR^2\to\bR$ a generic quasiperiodic function on the plane with 3 quasiperiods, namely $q=F_\psi$ with $F$
  a generic element of $C^\infty(T^3)$ and $\psi$ a fully irrational element of $A_{3,2}$, and denote by $B_\psi\in\RPt$
  the direction perpendicular to the plane $\psi(\bR^2)$ in $\bR^3$ and by $\psi_a$, $a\in\Rt$, any sibling of $\psi$.
  Set $q_a=F_{\psi_a}$. Then:
  \begin{enumerate}
  \item all connected components of the  level sets $q_a=c$ are closed when $c\not\in[\low{F}(B_\psi),\upp{F}(B_\psi)]$;
  \item if $B_\psi\in\cD_F$, then all level sets $q_a=c$, with $c\in[\low{F}(B_\psi),\upp{F}(B_\psi)]$, contain
    non-singular open components and, if $B_\psi$ is fully irrational, all of them are strongly asymptotic to the direction
    $B_\psi\times\ell_F(B_\psi)$;
  \item if $B_\psi\in\cE_F$, then the level set  $q_a=e$, with $e=\low{F}(B)=\upp{F}(B)$, has open connected component(s), 
    none of which strongly asymptotic to a straight line.
  \end{enumerate}
\end{theorem}
\begin{remark}
  When $\psi$ is not fully irrational, namely when its image contains some rational direction,
  open (periodic) orbits might arise for a larger closed connected interval
  of values of $q_a$ but such orbits are {\sl unstable}, namely they disappear for a generic perturbation of $\psi$.
\end{remark}
One of the main points of Theorem~1 is the discovery of the hidden topological first integral
$\ell_F(B_\psi)$, namely a triple of coprime integers $(m_1,m_2,m_3)$ locally constant with
respect to $B_\psi$, that dictates the asymptotics of open level sets when $B_\psi\in\cD_F$
in such a way that they behave just as if $F(x)=m_1 x^1+m_2 x^2+m_2 x^3$.
In the most interesting cases, the dependence of $\ell_F$ on $B_\psi$ is of fractal nature
(see Fig.~\ref{fig:cos3D} for some concrete example).

There are still three intertwined fundamental questions left unanswered here:
\renewcommand\labelitemi{$\spadesuit$}
\begin{itemize}
  \item Is $\cE_F$ a zero-measure set for a generic $F$?
  \item If so, does $\cE_F$ have non-integer Hausdorff dimension?
  \item What is the geometry of the level sets open components when $B_\psi\in\cE_F$?
\end{itemize}
\begin{conjecture}[Novikov~\cite{NM04}]
  For a generic  $F\in C^\infty(\bT^3)$, the set $\cE_F$ has zero measure and Hausdorff dimension strictly between 1 and 2.
\end{conjecture}
In the only particular example where it was possible to find an analytical
expression for the map $\ell_F$, introduced by the first author and Dynnikov, it was proved~\cite{DD09}
that $\cE_F$ was indeed a null set; later Avila, Hubert and Skripchenko~\cite{AHS16b} showed that its
Hausdorff dimension is indeed strictly smaller than 2. About the last question, Skripchenko
and Dynnikov built examples of $F$ such that each $F_{\psi_a}$ has a unique open level set~\cite{Skr13}
and such that there are infinitely many~\cite{DS15}. It is not clear yet what is the generic situation.

In case of 4 quasiperiods the situation is much more unclear. The only results are due to Novikov
and Dynnikov~\cite{Nov99,DN05} and cover only the {\sl close-to-rational} case:
\begin{definition}
  Let $G_{4,2}$ be the Grassmannian space of all 2-dimensional linear subspaces of $\bR^4$.
  Given a $B\in G_{4,2}$ and a direction $\ell\in(\RPf)^*$, let $\Pi$ be a 2-plane with $[\Pi]=B$
  and $L$ a covector with $[L]=\ell$. We denote by $B\times\ell\in\RPf$ the direction of the line
  obtained by intersecting $\Pi$ with the hyperplane $\ker L\subset\bR^4$.
\end{definition}
%
%
\begin{theorem}[Dynnikov, Novikov~\cite{DN05}]
  \label{lemma:4qp}
  For every generic function $F\in C^\infty(\bT^4)$, there is an open dense set $\cD_F\subset G_{4,2}$
  and a locally constant function $\ell_F:\cD_f\to\QPT$ such that
  all non-singular open sections of any level surface of $F$ with any 2-plane $\psi(\Rt)$,
  with $B_\psi\in\cD_F$, are strongly asymptotic to a straight line with direction $B\times\ell_F(B_\psi)$. 
  \end{theorem}
As a corollary of this theorem, we get the following claim about the level sets of QP functions:
\begin{theorem}
  \label{thm:4qp}
  Let $q:\Rt\to\bR$ be a quasiperiodic function on the plane with 4 quasiperiods such that $q=F_\psi$, where
  $F\in C^\infty(\bT^4)$ is generic and $B_\psi\in\cD_F\subset G_{4,2}$, and let $q_a=F_{\psi_a}$,
  where $\psi_a$, $a\in\bR^2$, be any sibling of  $\psi$. Then 
  all non-singular open components of all level sets $q_a=c$ are strongly asymptotic to the direction
  $B_\psi\times\ell_F(B_\psi)$.
\end{theorem}
%
%
%
\section{QP functions in electron transport phenomena}
\label{sec:Mal}
To describe the applications of the Novikov problem in the 
transport phenomena in normal metals we have to start with a
description of electron states in a crystal lattice, defined by
bounded solutions of the Shr\"odinger equation
\begin{equation}
  \label{ShredingerEquation}
  - \, {\hbar^{2} \over 2 m} \,\, \Delta \psi  \,\,\, + \,\,\, 
  U (x, y, z) \, \psi \,\,\, = \,\,\, \varepsilon \, \psi  \,\,\,  
\end{equation}
The potential $\,\, U ({\bf x}) \, = \, U (x, y, z) \,\, $ represents
a periodic function in $\, \mathbb{R}^{3} \, $ with 
three different periods
$\, {\bf l}_{1}$, $\, {\bf l}_{2}$, $\, {\bf l}_{3}$:
$$U ({\bf x} + {\bf l}_{1}) \,\,\, \equiv \,\,\, 
U ({\bf x} + {\bf l}_{2}) \,\,\, \equiv \,\,\,
U ({\bf x} + {\bf l}_{3}) \,\,\, \equiv \,\,\, U ({\bf x}) 
\,\,\, , $$
which define the crystal lattice $\, L \, $ of a metal. 

 The basis physical solutions of the equation (\ref{ShredingerEquation})
can be chosen in the form of the Bloch functions 
$\, \psi_{\bf p} ({\bf x}) $, satisfying the conditions
$$\psi_{\bf p} ({\bf x} + {\bf l}_{1}) \,\,\, \equiv \,\,\,
e^{i ({\bf p} , {\bf l}_{1}) / \hbar} \,\,
\psi_{\bf p} ({\bf x})  \,\,\, ,  \quad
\psi_{\bf p} ({\bf x} + {\bf l}_{2}) \,\,\, \equiv \,\,\,
e^{i ({\bf p} , {\bf l}_{2}) / \hbar} \,\,
\psi_{\bf p} ({\bf x})  \,\,\, ,   $$
$$\psi_{\bf p} ({\bf x} + {\bf l}_{3}) \,\,\, \equiv \,\,\,
e^{i ({\bf p} , {\bf l}_{3}) / \hbar} \,\,
\psi_{\bf p} ({\bf r}) $$

 The real vector $\, {\bf p} \, = \, (p_{1}, p_{2}, p_{3}) \, $
represents the quasimomentum of an electron state and is defined
in fact modulo the vectors
\begin{equation}
\label{ReciprLattice}
m_{1} \, {\bf a}_{1} \,\,\, + \,\,\, m_{2} \, {\bf a}_{2} 
\,\,\, + \,\,\, m_{3} \, {\bf a}_{3} \,\,\, ,  
\quad  m_{1}, m_{2}, m_{3} \, \in \, \mathbb{Z} \,\,\, , 
\end{equation}
where the vectors 
$\, {\bf a}_{1} $,  $\, {\bf a}_{2} $, $\, {\bf a}_{3} \, $
are defined by the relations
$${\bf a}_{1} \,\,\, = \,\,\, 2 \pi \hbar \,\,
{{\bf l}_{2} \times {\bf l}_{3} \over 
({\bf l}_{1}, {\bf l}_{2}, {\bf l}_{3})}
\,\,\, ,  \quad
{\bf a}_{2} \,\,\, = \,\,\, 2 \pi \hbar \,\,
{{\bf l}_{3} \times {\bf l}_{1} \over 
({\bf l}_{1}, {\bf l}_{2}, {\bf l}_{3})}
\,\,\, ,  \quad
{\bf a}_{3} \,\,\, = \,\,\, 2 \pi \hbar \,\,
{{\bf l}_{1} \times {\bf l}_{2} \over 
({\bf l}_{1}, {\bf l}_{2}, {\bf l}_{3})} $$

 The vectors $\, {\bf a}_{1} $,  $\, {\bf a}_{2} $, $\, {\bf a}_{3} \, $
give a basis of the reciprocal lattice $\, L^{*} \, $ of a crystal, 
conjugate to the direct lattice $\, L \, $. In general, the full space
of physical solutions of (\ref{ShredingerEquation}) consists of an
infinite number of ``energy bands'' where the dependence of the parameter 
$\, \varepsilon \, $ on the value of $\, {\bf p} \, $ is given by some
three-periodic smooth functions $\, \varepsilon_{s} ({\bf p}) $:
$$\varepsilon_{s} ({\bf p} + {\bf a}_{1}) \,\,\, \equiv \,\,\,
\varepsilon_{s} ({\bf p} + {\bf a}_{2}) \,\,\, \equiv \,\,\,
\varepsilon_{s} ({\bf p} + {\bf a}_{3}) \,\,\, \equiv \,\,\, 
\varepsilon_{s} ({\bf p}) $$

 Thus, the complete set of parameters specifying single-electron states
in a crystal includes the number of the conduction band $\, s \, $, the
quasimomentum value $\, {\bf p} \, $, and the spin variable 
$\, \sigma \, $. The last variable will in fact not be important 
in our considerations, so we will omit it in our further constructions.

 For a fixed energy band any two values of the quasimomentum that
differ by any reciprocal lattice vector define the same physical
electron state. As a result, we can actually claim that the space
of electron states for a fixed energy band represents a
three-dimensional torus $\, \mathbb{T}^{3}$:
$$\mathbb{T}^{3} \,\,\, = \,\,\, \mathbb{R}^{3} / L^{*} \,\,\, , $$
given by the factorization of the $\, {\bf p}$-space over the
reciprocal lattice vectors. In the same way, every dispersion
relation $\, \varepsilon_{s} ({\bf p}) \, $ can be considered as a 
smooth function on $\, \mathbb{T}^{3} \, $ instead of the full
Euclidean $\, {\bf p}$-space $\, \mathbb{R}^{3} \, $. Every
function $\, \varepsilon_{s} ({\bf p}) \, $ is naturally bounded
by its minimal and maximal values
$$\varepsilon^{min}_{s} \,\,\, \leq \,\,\, \varepsilon_{s} ({\bf p}) 
\,\,\, \leq \,\,\, \varepsilon^{max}_{s} \,\,\, , $$
which define the boundaries of the corresponding energy band.
Let us also note here that in the three-dimensional case the
intervals $\, [\varepsilon^{min}_{s} , \varepsilon^{max}_{s}] \, $
can in general overlap with each other, so maybe it would be more
rigorous to talk about different branches of the electron energy 
spectrum in a crystal.  

 Practically in any metal, the electron gas is highly degenerate 
and it can be assumed that all the electron states with energies 
below a certain value $\, \varepsilon_{F} \, $ (the Fermi energy) are
occupied, while states with energies greater than the Fermi energy 
are empty. In the general case, we have here a certain finite number 
of completely filled energy bands, a finite number of partially filled
bands (conduction bands), and an infinite number of empty energy bands.
The full Fermi surface of a metal is given by the union of the 
surfaces 
\begin{equation}
\label{FermiSurface}
\varepsilon_{s} ({\bf p}) \,\,\, = \,\,\, \varepsilon_{F}
\end{equation}
for all partially filled energy bands and represents a 
3-periodic smooth surface in the $\, {\bf p}$-space.

 We would like to specially note here that we do not require that the 
Fermi surface consists of only one connected component. For us it is 
important, however, that the different connected components of the Fermi
surface do not intersect each other. We note here that the latter 
property is satisfied, as a rule, also in the case when the Fermi 
surface is determined by several dispersion relations.

 The form of the dispersion relation $\, \varepsilon ({\bf p}) \, $
is very important for many quantum processes in crystals and, in
particular, in normal metals. For us, the processes associated 
with transport phenomena in metals, for which the dynamics of quantum 
electron states in the presence of external electric and magnetic 
fields, will play a decisive role. It can immediately be noted that, 
since the magnitude of external electric and magnetic fields is much 
smaller than the magnitude of the intracrystalline fields, such dynamics 
are well described by the adiabatic approximation for the evolution of 
the functions $\, \psi_{\bf p} ({\bf r}) $, which can be written in the 
form of a dynamical system determining the evolution of the values 
of the quasimomentum $\, {\bf p} \, $. Thus, in the presence of constant 
external electric and magnetic fields, the corresponding system can be 
written in the form (see e.g. \cite{Abr88,Kit63,Zim72})
\begin{equation}
  \label{AdiabaticSystem}
        {\dot {\bf p}} \,\,\,\,\, = \,\,\,\,\, {e \over c} \,
        \left[ {\bf v}_{gr} \times {\bf B} \right] 
        \,\, + \,\, e \, {\bf E} \,\,\,\,\, \equiv \,\,\,\,\, {e \over c} \, 
        \left[ \nabla \varepsilon ({\bf p}) \times {\bf B} \right]
        \,\, + \,\, e \, {\bf E} 
\end{equation}
 The electron transport properties are determined in the main order 
by the properties of solutions of the kinetic equation for the 
one-particle distribution function $\, f ({\bf p}, t) \, $, 
which can be written in the general case in the form
\begin{equation}
  \label{KinEq}
  f_{t} \,\,\, + \,\,\, {e \over c} \, \sum_{k=1}^{3} \,
  \left[ \nabla \varepsilon ({\bf p}) \times {\bf B} \right]^{k} \,\,
       {\partial f \over \partial p^{k}} \,\,\, + \,\,\, e \, \sum_{k=1}^{3} \,
       E^{k} \, {\partial f \over \partial p^{k}}
       \,\,\,\, =  \,\,\,\, I [f] ({\bf p}, \, t)
\end{equation}
 The functional $\, I [f] ({\bf p}, \, t) \, $ is the collision 
integral, which in the general case determines the relaxation of the 
perturbations of the function $\, f ({\bf p}, t) \, $ to its
temperature-equilibrium values
\begin{equation}
  \label{DistrFunct}
  f_{0} ({\bf p}) \,\,\, = \,\,\, 
  {1 \over e^{(\varepsilon ({\bf p}) - \varepsilon_{F}) / T} \, + \, 1 }
\end{equation}
 Quite often, all the necessary properties of the solutions 
of (\ref{KinEq}) can be obtained by introducing a certain typical 
relaxation time of the function $\, f ({\bf p}, t) \, $ 
to its equilibrium values (the mean free electron time) 
$\, \tau \, $ and replacing the collision integral by the value
$$- \,\, \left. \Big( f ({\bf p}, t) - f_{0} ({\bf p}) \Big)
\right/ \tau $$

 When calculating the electronic transport properties in metals 
(such as electrical conductivity or electron thermal conductivity), 
the most interesting is usually the response of the system to the 
application of, say, an electric field (or a temperature gradient)
in the linear approximation in the value of $\, {\bf E} \, $.
From this point of view, the electric field in the equations 
(\ref{AdiabaticSystem}) can be regarded as a small correction 
to the system
\begin{equation}
  \label{MagnFieldSystem}
        {\dot {\bf p}} \,\,\,\,\, = \,\,\,\,\, {e \over c} \, 
        \left[ \nabla \varepsilon ({\bf p}) \times {\bf B} \right] \,\,\, ,
\end{equation}
determining the evolution of electron states in the presence
of a constant magnetic field. The geometry of the trajectories 
of the system (\ref{MagnFieldSystem}) plays an important role 
for the corresponding electronic properties of a metal in the 
region $\, \omega_{B} \tau \, \gg \, 1 \, $, where the cyclotron
frequency $\, \omega_{B} \, $ is defined by the relation
$\, \omega_{B} \, = \, e B / m^{*} c \, $. Let us note also 
here that the value of the effective electron mass $\, m^{*} \, $
in a metal can, in general, differ noticeably from the free 
electron mass $\, m \, $. 

 In the semiclassical approximation one can also consider the motion 
of electron wave packets in the coordinate space, which is given by 
the relations
$${\dot {\bf x}} \,\,\,\,\, = \,\,\,\,\, {\bf v}_{gr} ({\bf p})
\,\,\,\,\, \equiv \,\,\,\,\, \nabla \varepsilon \, ({\bf p}) $$

 It is not difficult to see here that the corresponding electron 
trajectories in $\, {\bf x}$-space are closely connected with 
the trajectories given by the system (\ref{MagnFieldSystem}).
In particular, the projections of the electron trajectories in 
the $\, {\bf x}$-space onto a plane orthogonal to $\, {\bf B} \, $ 
are obtained from the trajectories of the system (\ref{MagnFieldSystem})
in $\, {\bf p}$-space by rotation through an angle of $90^{\circ}$ in 
the same plane. The latter circumstance clarifies the role of the shape 
of trajectories of the system (\ref{MagnFieldSystem}) for electron 
transport phenomena and it can be also seen that the parameter 
$\, \omega_{B} \tau \, $ determines the average length of the electron 
motion along the trajectory between two acts of scattering by an impurity.
The condition $\, \omega_{B} \tau \, \gg \, 1 \, $ then leads to 
the manifestation of the features of the global geometry of the 
trajectories of system (\ref{MagnFieldSystem}) for phenomena of this type.

As pointed out in the introduction, system (\ref{MagnFieldSystem}) is equivalent
to the Hamiltonian dynamical system in $\bT^3$ given by
$$\dot p_i=\{p_i,\varepsilon(\p)\}_B\,,\; i=1,2,3\,,\;\hbox{ with }\;\{p_i,p_j\}_B = \epsilon_{ijk}B^k\,.$$
A Poisson bracket in odd dimension is necessarily degenerate and, in fact, $\{,\}_B$ has the (multivalued!)
Casimir $b(\p) = B^ip_i$. On each of the Casimir's leaves, namely on each of the projections into $\bT^3$
of the planes $\psi_a:\bR^2\to\bR^3$ perpendicular to $\B$, the system is non-degenerate and equivalent
to the Hamiltonian system with QP Hamiltonian $\varepsilon_{\psi_a}$. Hence, since we are in dimension 2,
the orbits of the solutions of~(\ref{MagnFieldSystem}) are just the (projection into $\bT^3$ of) level
sets of $\varepsilon_{\psi_a}$, whose structure has been summarized in Theorem~1.

\begin{figure}[t]
  \begin{center}
    \includegraphics[width=0.9\linewidth]{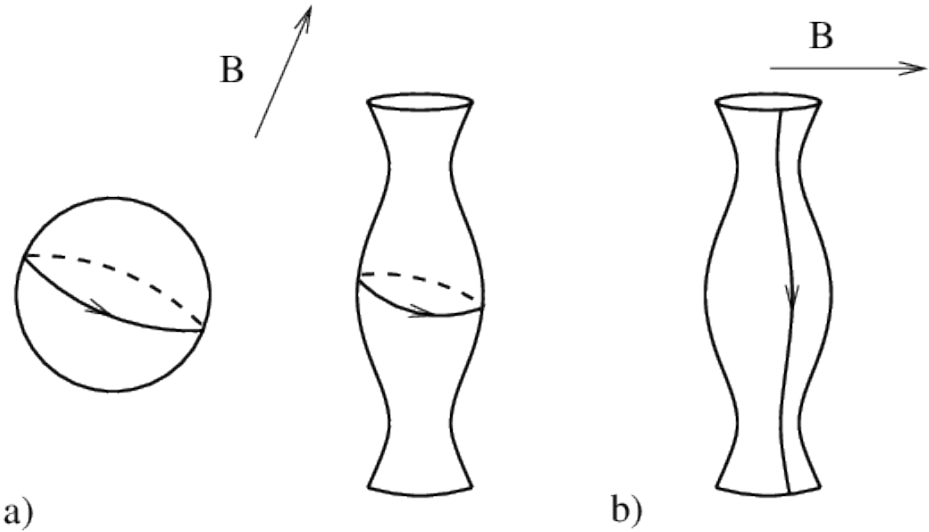}
  \end{center}
  \caption{The Fermi surfaces containing only closed trajectories (a)
    and the Fermi surface containing open periodic trajectories at special 
    directions of $\B$ (b).}
  \label{ClosedAndPeriodic}
\end{figure}

The study of the question of the influence of the geometry of trajectories of the system~(\ref{MagnFieldSystem})
on the behavior of electron transport phenomena was started in the school of  I.M. Lifshitz in the 1950s
(see \cite{LAK56,LP59,LP60,LK60,LK63,LK66,LAK73}).
Thus, in the work~\cite{LAK73} it was first shown that the behavior of the electric conductivity tensor of a metal in strong magnetic fields 
is significantly different in the cases when the Fermi surface contains only closed trajectories and when open periodic trajectories appear on it
(Fig.~\ref{ClosedAndPeriodic}). Let us always assume here that the coordinate system in the  $\, {\bf x}$-space is chosen in such a way that the 
$z$-axis coincides with the direction of the magnetic field. In addition, let us also assume that the direction of the $x$-axis 
in the second case coincides with the mean direction of the periodic open trajectories in the $\, {\bf p}$-space (note here that the mean direction 
of the projection of the corresponding trajectories onto the plane orthogonal to $\B$ in the $\, {\bf x}$-space coincides 
with the $y$-axis in this case). Then, according to~\cite{LAK57}, the analysis of the equation~(\ref{KinEq}) gives the following results for  
the asymptotic behavior of the conductivity tensor in the two cases above 
\begin{equation}
  \label{Closed}
  \sigma^{ik} \,\,\,\, \simeq \,\,\,\,
        {n e^{2} \tau \over m^{*}} \, \left(
        \begin{array}{ccc}
          ( \omega_{B} \tau )^{-2}  &  ( \omega_{B} \tau )^{-1}  &
          ( \omega_{B} \tau )^{-1}  \cr
          ( \omega_{B} \tau )^{-1}  &  ( \omega_{B} \tau )^{-2}  &
          ( \omega_{B} \tau )^{-1}  \cr
          ( \omega_{B} \tau )^{-1}  &  ( \omega_{B} \tau )^{-1}  &  *
        \end{array}  \right)  ,   \quad \quad
        \omega_{B} \tau \,\, \rightarrow \,\, \infty 
\end{equation}
(closed trajectories),
\begin{equation}
  \label{OpenPeriodic}
  \sigma^{ik} \,\,\,\, \simeq \,\,\,\,
        {n e^{2} \tau \over m^{*}} \, \left(
        \begin{array}{ccc}
          ( \omega_{B} \tau )^{-2}  &  ( \omega_{B} \tau )^{-1}  &
          ( \omega_{B} \tau )^{-1}  \cr
          ( \omega_{B} \tau )^{-1}  &  *  &  *  \cr
          ( \omega_{B} \tau )^{-1}  &  *  &  *
        \end{array}  \right)  ,   \quad \quad
        \omega_{B} \tau \,\, \rightarrow \,\, \infty 
\end{equation}
(open periodic trajectories).

We note here that the relations (\ref{Closed}) and (\ref{OpenPeriodic}) 
should be understood only as asymptotic expressions and may contain 
additional dimensionless constants for each of the components
$\, \sigma^{ik} \, $. We also use here the notation $\, * \, $ for 
arbitrary dimensionless constants of the order of unity.

 It is easy to see that the main difference in the conductivity behavior 
in the cases considered above is the strong anisotropy of the conductivity 
in the plane, orthogonal to $\B$, observed in the second case.
This property is a direct consequence of the special form of the corresponding 
electron trajectories and makes it possible to measure the mean direction 
of the periodic open trajectories in the $\, {\bf p}$-space.

 In the works \cite{LP59,LP60}, open trajectories of a more general 
type on Fermi surfaces of different shapes were considered. Let us say, that
the trajectories, considered in~\cite{LP59,LP60}, are not periodic
in general, but also have a mean direction in the plane orthogonal to 
$\B$. As a result, the conductivity behavior also exhibits 
strong anisotropic properties in strong magnetic fields in the presence 
of trajectories of this type on the Fermi surface. The works
\cite{LK60,LK63,LK66}, as well as the book~\cite{LAK73}, provide 
a broad overview of the issues related to the electronic properties of metals, 
and in particular the issues related to transport phenomena in strong 
magnetic fields examined during that period. We would also like to give 
here a reference to the work \cite{KP02} in which a return 
to this range of issues is made after a considerable time, and containing 
also aspects that arose in the later period. 

 As we have said above, the problem of the complete classification of 
possible types of trajectories of the system (\ref{MagnFieldSystem})
was set by S.P. Novikov in the early 1980s and has now been studied 
with sufficient completeness, allowing to describe all essentially 
different types of open electron trajectories. In this chapter we will 
focus on the most significant physical results arising from the mathematical 
description of the trajectories of system (\ref{MagnFieldSystem}),
obtained in the recent decades.

 As we noted in the previous chapter, the most significant part in the 
classification of open trajectories of system (\ref{MagnFieldSystem}) is 
the description of stable open trajectories obtained in the works of 
A.V. Zorich and I.A. Dynnikov. We shall try to describe here the most 
interesting physical consequences arising when such trajectories appear 
on the Fermi surface. 

\begin{figure}[t]
\begin{center}
\includegraphics[width=0.9\linewidth]{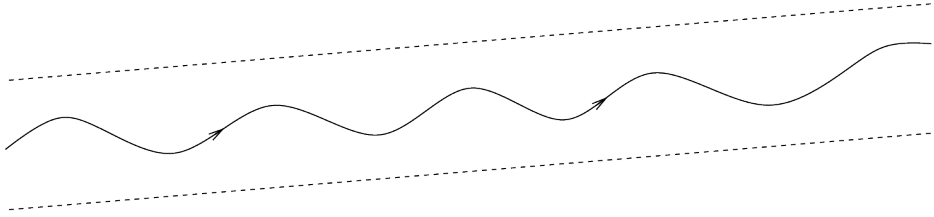}
\end{center}
\caption{A stable open trajectory in the plane orthogonal to $\B$ in the $\, {\bf p}$-space.}
\label{StableTraject}
\end{figure}

Since the orbits of the solutions of (\ref{MagnFieldSystem}) are the level
sets of all siblings $\varepsilon_{\psi_a}$ of the quasiperiodic function in
two variables with three quasiperiods $\varepsilon_\psi$,
by Theorem~1 such trajectories always possess the following two
remarkable properties:
\begin{enumerate}
  \item Any stable open trajectory of system (\ref{MagnFieldSystem}) in 
    the $\, {\bf p}$-space lies in a straight strip of finite width in 
    a plane $\psi_a$ (Fig. \ref{StableTraject});
  \item For a fixed direction $\B$, all stable open trajectories
    in the $\, {\bf p}$-space have the same mean direction, given
    by $\B\times \ell_\varepsilon({\bf B})$.
\end{enumerate}

As pointed out in~\cite{NM96}, the presence of stable open trajectories on the Fermi surface always entails
a strong anisotropy of the conductivity tensor in the plane, orthogonal to $\B$, in the limit 
$\, \omega_{B} \tau \, \rightarrow \, \infty \, $. Because of this, the topological quantum first integral
$\ell_\varepsilon(\B)$ is {\sl observable} experimentally. We call {\sl Stability Zones} the sets
$\cD_{l,\varepsilon}$ defined by $\ell_\varepsilon(\B)=l$, so that for every $\B\in\cD_{l,\varepsilon}$
all non-singular open orbits are strongly asymptotic to $\B\times l$.

Both the topological quantum first integral and the geometry of the Stability Zones contain important
information about the electron spectrum in a crystal that is directly related to the determination 
of parameters of this spectrum in real materials. At the same time, both theoretical and experimental
determination of the exact boundaries of the Stability Zones for a given dispersion relation represents
a non-trivial problem that requires the use of rather special methods. As an example of a theoretical 
determination of the boundaries of the Stability Zones, we can cite the work 
\cite{DeL05b}, where such calculations were performed for a number of 
analytical dispersion relations that arise in real crystals. As can be seen 
from the work~\cite{DeL05b}, an accurate calculation of the structure of 
the Stability Zones on the angular diagram requires the development of both 
rather serious topological and computational methods. We hope, on the other hand, 
that the methods used in~\cite{DeL05b} must be applicable to a large number of
different examples of complex Fermi surfaces and will prove extremely useful in
determination the parameters of the dispersion relations in real materials.
It must also be said that the experimental determination of the structure 
of the Stability Zones in real materials also presents a special problem 
because of a rather complicated analytical behavior of conductivity near 
their boundaries (see, e.g.~\cite{Mal17b}). In particular, the exact experimental 
determination of the mathematical boundaries of the Stability Zones also requires, 
in addition to direct study of conductivity, special experimental techniques
(\cite{Mal17a}).

\vspace{1mm}

 Another very important achievement of mathematical research of the S.P. Novikov 
problem was the discovery of new, previously unknown, types of trajectories of 
system (\ref{MagnFieldSystem}), which have very complicated (chaotic) behavior. 
The first trajectories of this type were constructed at the beginning of Nineties
by S.P. Tsarev\footnote{Private communication}
for ``partially irrational'' directions of $\B$ and have an obvious 
chaotic behavior on the Fermi surface. At the same time, the behavior of the 
Tsarev trajectories in planes orthogonal to $\B$ resembles the behavior 
of stable open trajectories, in particular, they all have asymptotic directions 
in these planes (although they do not lie in straight strips of finite width). 
As a result, the behavior of the conductivity tensor in the presence of the Tsarev 
trajectories on the Fermi surface is also very similar to its behavior in 
the presence of stable open trajectories, in particular, it has a strong 
anisotropy in this case. As already mentioned, trajectories of Tsarev type 
can appear only for directions of the magnetic field of irrationality 2 
(the plane orthogonal to $\B$ contains a reciprocal lattice vector) 
and it can be shown (see~\cite{Dyn97}) that all chaotic trajectories arising 
for such directions of $\B$ have the properties described above.

\begin{figure}[t]
\begin{center}
\includegraphics[width=0.9\linewidth]{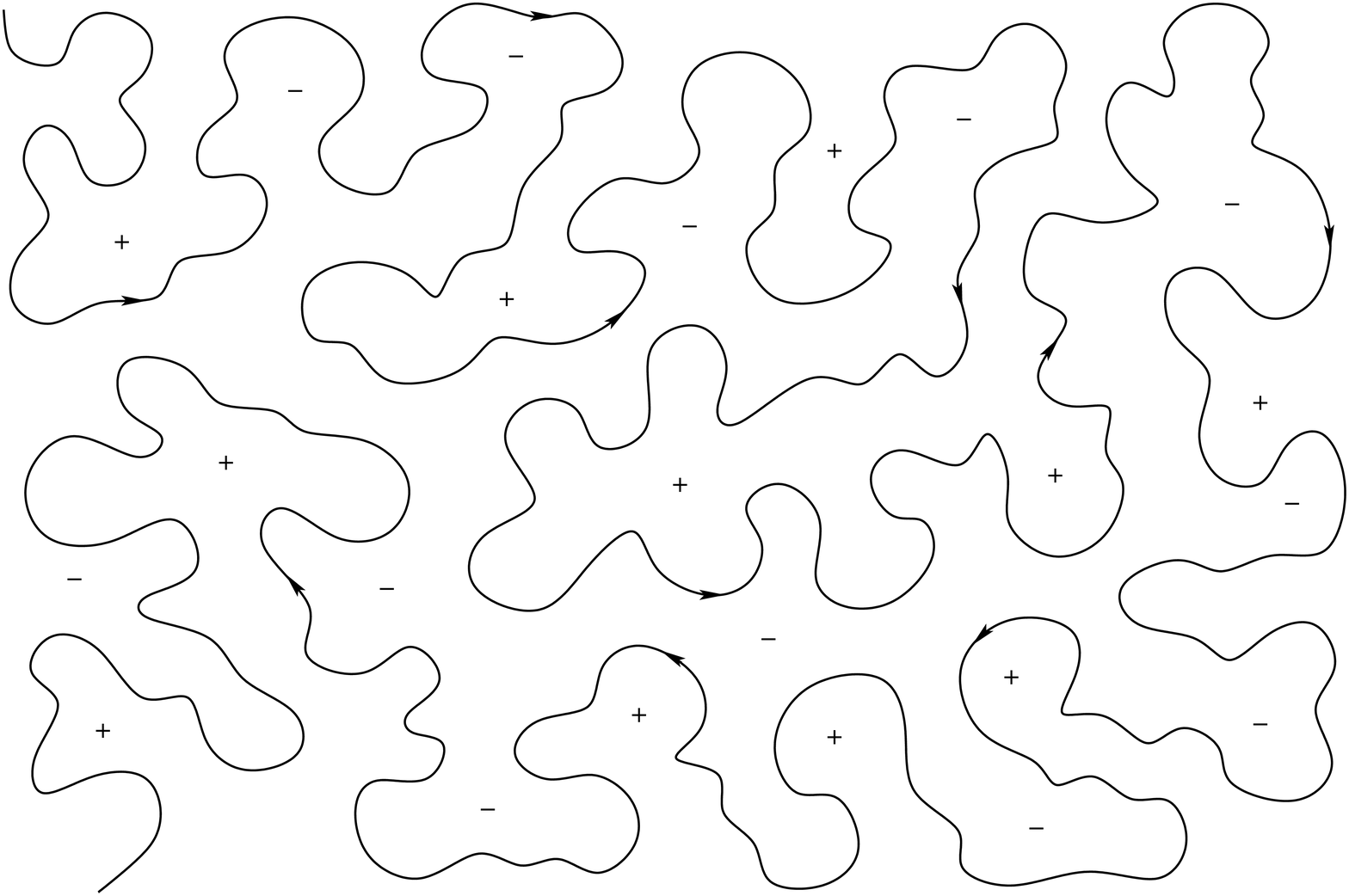}
\end{center}
\caption{The geometry of the Dynnikov chaotic trajectory in the plane
orthogonal to $\, {\bf B} $.}
\label{DynnTraject}
\end{figure}

 The first examples of chaotic trajectories for directions of $\B$ 
 of maximal irrationality were constructed by I.A. Dynnikov in the work~\cite{Dyn97}.
 Trajectories of this type have a strongly chaotic behavior 
both on the Fermi surface and in planes orthogonal to $\B$
(Fig.~\ref{DynnTraject}). As can be seen, the behavior of a Dynnikov trajectory 
in a plane orthogonal to $\B$ resembles in a certain sense the 
diffusion motion, which leads to the most complicated dependence of the 
conductivity on the value of $\, B \, $. 

 The most interesting moment in the behavior of conductivity 
in the presence of the Dynnikov trajectories is the blocking of the 
conductivity along the direction of  $\B$ in strong magnetic 
fields (\cite{Mal97}), such that the entire Fermi surface area covered by the 
corresponding chaotic trajectories does not contribute to the conductivity 
along $\B$ in the limit 
$\, \omega_{B} \tau \, \rightarrow \, \infty \, $.
As a result, for the corresponding directions on the angular diagram, 
rather sharp minima in conductivity along the direction of $\B$ 
should be observed in strong magnetic fields. 

 Another interesting feature of the conductivity behavior in the presence 
of the Dynnikov trajectories on the Fermi surface is the appearance of 
fractional powers of the parameter $\, \omega_{B} \tau \, $ in the dependence 
of the components of the conductivity tensor on the value of the magnetic 
field (\cite{Mal97}). It must be said that the analysis carried out for 
equation (\ref{KinEq}) in the presence of such trajectories actually used 
in this case an additional property (self-similarity) of trajectories constructed 
in~\cite{Dyn97}, which, generally speaking, is not observed in the general case 
for Dynnikov chaotic trajectories. Quite recently, however, it was possible 
to show that the appearance of fractional powers of a parameter 
$\, \omega_{B} \tau \, $ in the conductivity behavior in the presence 
of trajectories of this type is actually a common fact and is connected with 
the existence of the so-called Zorich - Kontsevich - Forni indices 
(see~\cite{Zor94,Zor96,Zor97,Zor97b,Zor99,Zor06}) 
describing the behavior of such trajectories on a large scale (\cite{NM18}).

 As mentioned in the previous section, the general properties of chaotic trajectories of 
Dynnikov type, as well as the properties of the set of directions of $\B$
at which such trajectories can be observed, are the subject of the most active research 
at the present time.
Let us also note here that, in spite of the 
fact that Dynnikov chaotic trajectories are not trajectories of general position, 
they may nevertheless be typical for Fermi surfaces of a certain type 
(see~\cite{Mal18}).

 Let us say, that, in addition to describing the new quantities and the new regimes 
observed in conductivity studies, a mathematical investigation of the Novikov problem 
actually made it possible to construct a complete classification of the possible types 
of conductivity behavior in strong magnetic fields, including all cases of both 
generic and non-generic position. Here we only point out that the most detailed 
mathematical consideration of the various situations possible for system 
(\ref{MagnFieldSystem}) is presented in the work~\cite{Dyn99}. We also note 
that a detailed exposition of the physical consequences of the classification 
obtained can be found in the works \cite{NM98,NM03a,NM04}. 

\vspace{1mm}

 We would also like to describe here another application of the Novikov problem 
related to transport phenomena in two-dimensional electron systems in a magnetic 
field, placed in an artificially created potential in the plane of the system.

 It is well known that two-dimensional electron systems are of great interest 
from the point of view of modern experimental and theoretical physics in connection 
with their very different applications in various fields. A special class of such 
systems are systems in an external magnetic field, which can be either parallel or
perpendicular to the plane of the system. We will consider here systems in 
a perpendicular magnetic field, placed in a quasiperiodic plane potential, which 
can be created using a variety of techniques. 
 
\begin{figure}[t]
\begin{center}
\includegraphics[width=0.9\linewidth]{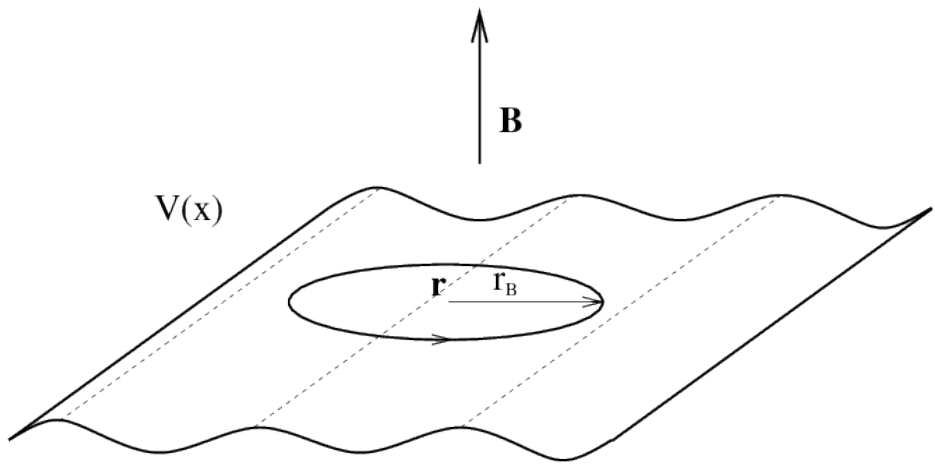}
\end{center}
\caption{The cyclotron electron orbit in the potential 
$\, V ({\bf x}) \, $.}
\label{CyclOrb}
\end{figure}

  We will be interested in the case when the behavior of electrons in a magnetic 
field can be described in the semiclassical approximation. In this case, we can 
imagine electrons moving along cyclotron orbits, the centers of which experience 
a slow drift under the influence of external fields (Fig. \ref{CyclOrb}). 
As an external field, we will consider a special external potential 
$\, V ({\bf x}) \, $ having quasiperiodic properties, which can also be 
considered as a model of a pseudo-random potential in the plane. We note here 
that many techniques for creating such potentials presuppose in reality the use 
of a superposition of periodic potentials with different periods (in the plane). 
It is easy to see that such techniques make it possible to obtain quasiperiodic 
potentials with any number of quasiperiods, which makes the problem under 
consideration quite relevant for such systems.

As it can be established (see e.g. \cite{Fer88,Bee89}), in the main order of the adiabatic approximation 
the centers of cyclotron orbits must drift along the level lines of a certain 
modified potential $\, {\bar V} ({\bf x}) \, $, obtained from the potential 
$\, V ({\bf x}) \, $ by averaging over the corresponding cyclotron electron 
orbits. It is also easy to see that the potential $\, {\bar V} ({\bf x}) \, $ 
has the same quasiperiodic properties as the potential $\, V ({\bf x}) \, $. 
As in the case of normal metals, the motion of electrons along the cyclotron 
orbits, as well as the drift of the cyclotron orbits in the potential, does not 
change the equilibrium statistical distribution of electrons in the two-dimensional 
system. At the same time, the peculiarities of the motion of the orbits along the 
level lines of the potential $\, {\bar V} ({\bf x}) \, $ have a significant effect 
on electronic transport phenomena that arise, say, in an electric field (electrical
conductivity), or in the presence of a temperature gradient (thermal conductivity). 
As a result, the description of the geometry of the level lines of a quasiperiodic 
potential $\, {\bar V} ({\bf x}) \, $ has the same significance in this problem 
as for electron transport phenomena in metals. 

 As we have seen above, the problem of the geometry of the level lines of 
quasiperiodic potentials on the plane can be considered at present solved 
for the case of three quasiperiods. In particular, we can also transfer here 
all the results concerning the description of electron transport phenomena 
determined by the geometry of the level lines of the potential 
$\, {\bar V} ({\bf x}) \, $. We note here that, unlike the case of normal 
metals, practically all the parameters of the problem are controlled in the 
described systems, so here it is possible to obtain any of the conductivity 
regimes described earlier. In particular, by changing the parameters of the 
potential, as well as the concentration of electrons in the system, we can 
easily achieve a regime of suppression of conductivity in the system 
(closed level lines), anisotropic conductivity regime (stable open level 
lines), or more complex regimes of diffuse conductivity (chaotic level lines).

 Among the cases with a larger number of quasiperiods, we should especially 
point the case of four quasiperiods, where deep topological results about 
regular level lines have now been obtained (\cite{Nov99,ND06}).
As for cases with five or more quasiperiods, at the moment the most probable 
is the prevalence here of complex chaotic regimes of behavior of the level 
lines of the corresponding potentials.
\section{Experimental and numerical study of level sets of QP functions on $\bR^2$ with 3 quasiperiods}
No algorithm is known to write an analytical or approximate perturbative expression
of the set $\cD_{F}$ and of the functions
$\low{F},\upp{F}$ relative to a general function $F\in C^\infty(\bT^3)$. In fact, an analytical
description for non-trivial $\cD_F$ and $\ell_F$ has been found only in case of a very simple
piecewise linear function~\cite{DD09}. Numerical methods are therefore necessary in order to
get some intuition on the nature of such sets and maps and in order to predict theoretically
from first principles the physical behavior of systems involving QP functions.
\begin{figure}
  \centering
  \includegraphics[width=8cm]{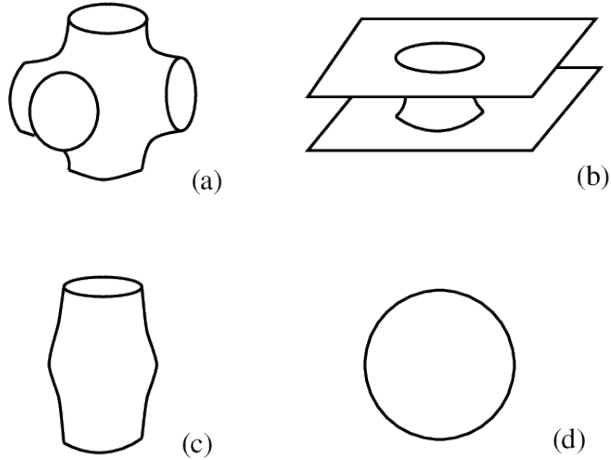}
  \caption{%
    \footnotesize
    Examples of surfaces embedded in $\bT^3$ with rank respectively 3,2,1,0.
  }
  \label{fig:rk}
\end{figure}

In order to illustrate the numerical algorithm we used so far to explore this problem
we need first to describe its {\sl extrinsic} geometry (e.g. see~\cite{Nov95}). In particular,
rather than considering the foliation induced on a plane $\psi:\bR^2\to\bR^3$
by the level sets of a periodic function $F$, we focus instead on the foliation induced on a
triply periodic surface $F_c=\{F=c\}\subset\bT^3$ by the bundle of (projections into $\bT^3$ of)
planes of all siblings of $\psi$.

In the generic case, $\psi$ is fully irrational and so the planar sections by the $\psi_a$
will cut $F_c$ in some finite number of open cylinders $C_i$ foliated by compact orbits and some
finite number of closed components $N_j$ equal to the closure of any open orbits in it.
Each of these $N_j$ has a boundary which is the union of finitely many loops homotopic to zero,
all of them lower or upper bases of some of the cylinders $C_i$. Since all the boundary components
are homotopic to zero, it makes sense to define the genus of the $N_j$ as the genus of the surfaces without
boundary $\tilde N_j$ obtained from the corresponding $N_j$ quotienting to a single point.
\begin{definition}
  Given a surface $M_g$ of genus $g$ and an embedding $i:M_g\to\bT^3$, we call {\sl rank}
  of the embedding the rank of the induced ring homomorphism $i_*: H_1(M_g,\bZ)\simeq\bZ^{2g}\to H_1(\bT^3,\bZ)\simeq\bZ^3$.
\end{definition}
Intuitively, if $M_g$ is embedded with rank $r$ then $\pi^{-1}(M_g)$ lies inside a
finite-width neighborhood of a $r$-dimensional linear subspace of $\bR^3$. In particular,
no open orbits can arise for generic $\psi$ when $r=0,1$ and the problem is trivial
when $r=2$, so we will assume from now on that $r=3$ (e.g. see Fig.~\ref{fig:rk}).
Two fundamental observation in elementary topology pointed out by Dynnikov
lead to the definition of $\ell_F$, namely:
\begin{theorem}[Dynnikov~\cite{Dyn96}]
  If a component $N_j$ contains an orbit whose counterimage in the universal covering
  is strongly asymptotic to a straight line, then $\tilde N_j$ must have genus 1 and must be embedded
  in $\bT^3$ with rank 2, namely $[\tilde N_j]$ is an indivisible integer 2-cycle in $H_2(\bT^3,\bZ)\simeq\bZ^3$.
  Moreover, in this case for all other components $N_{j'}$ of $F_c$ and of any other level surface $F_{c'}$
  hold same properties and all of them represent, modulo sign, the same 2-cycle $\l=[\tilde N_j]$.
\end{theorem}
Equivalently, by filling up with planar discs all holes of the counterimage of any one of these components
in the universal covering we get a {\sl warped plane}, namely an embedding $\bR^2\to\bR^3$
whose image lies in a finite-width region between two planes perpendicular to $\l$ (thought as a
direction in $\bR^3$).
This integral direction $\l$ is indeed the value of $\ell_F$ at $\B_\psi$.
\begin{figure}
  \centering
  \includegraphics[width=12cm]{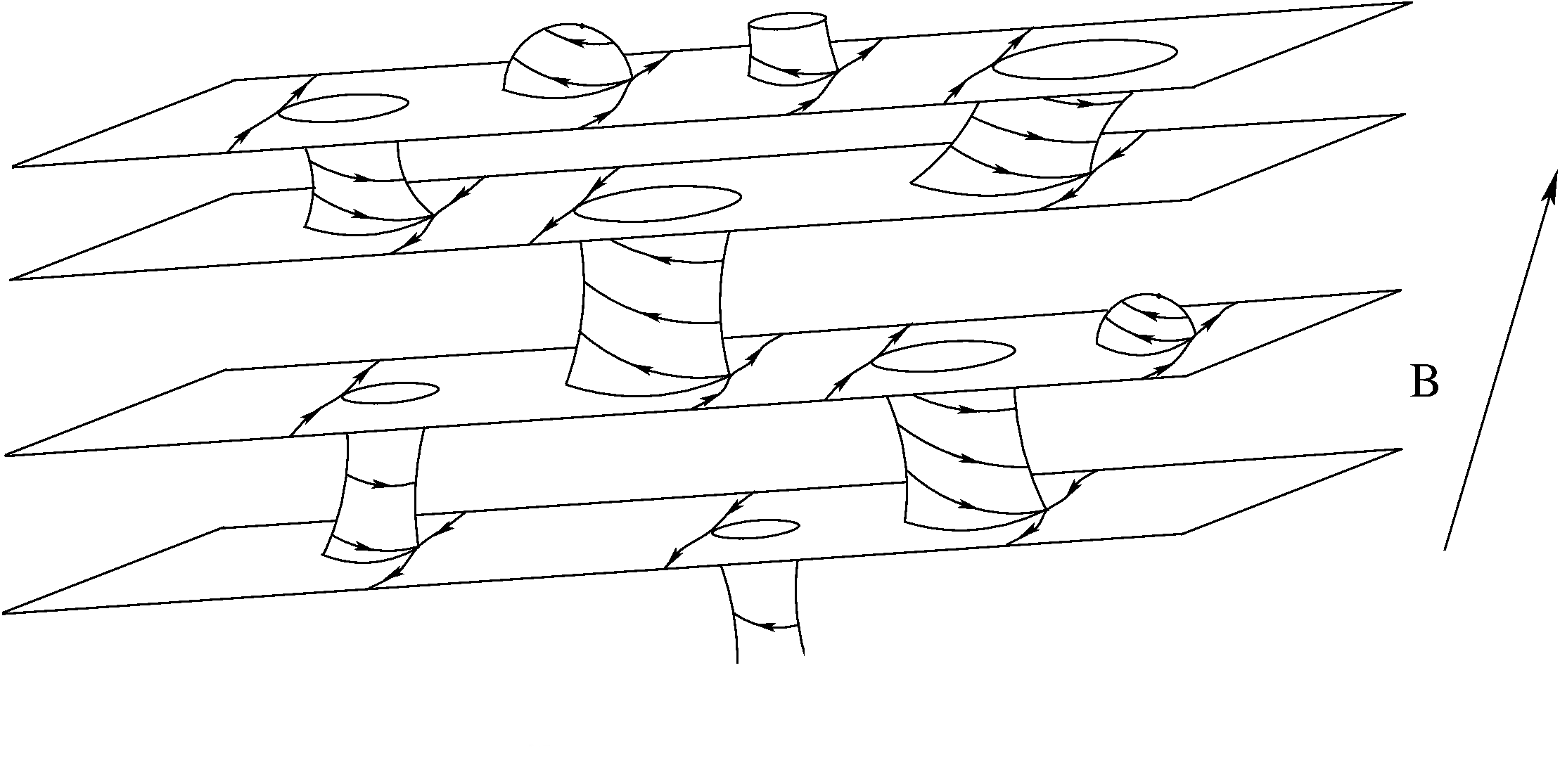}
  \caption{%
    \footnotesize
    Typical structure of the sections of a periodic surface by a bundle of
    parallel planes: in this case the surface gets subdivided into four $N_j$ components
    ( {\sl warped planes}) filled by open orbits, six cylinders filled by closed orbits and
    a few caps also filled by closed orbits.
  }
\label{fig:fol}
\end{figure}

Recall that in the homology of a manifold $N^n$ it is defined an intersection product
$\cap:H_i(N^n)\times H_{n-j}(N^n)\to \bZ$ defined so that $[a]\cap[b]$ is the signed
number of intersection of any two representatives of the cycles $a$ and $b$ transversal to each other.
Since two leaves of the same foliation cannot intersect each other, the non-trivial loops contained in every
genus-1 rank-2 component of $M_g$ do not intersect any of the trivial (in $\bT^3$) loops
in the cylinders $C_i$.
In order to find $\ell_F(\B_\psi)$, therefore, it is enough to find the rank-2 sublattice of
$H_1(\bT^3,\bZ)$ of all 1-cycles such that their homology intersection product with
$\ell_F(\B_\psi)$ is 0. This can be accomplished by selecting a canonical basis $\{e_i,f_j\}$
of $H_1(M_g,\bZ)$, namely so that $e_i\cap f_j=\delta_{ij}$, evaluating the
homology classes $c_i$ corresponding to each cylinder of compact orbits and
finding the symplectic orthogonal of the span of the $c_i$, namely the sublattice
$K\subset H_1(M_g,\bZ)$ of the homology classes whose intersection product
with all $c_i$ is zero. The push-forward $i_*K\subset H_1(\bT^3,\bZ)$ is
the rank-2 sublattice we were looking for.
%

The first (semi-analytic) study of a concrete case of QP function was done by
Dynnikov~\cite{Dyn96} for
$$
\fc(x,y,z)=\cos(2\pi x)+\cos(2\pi y)+\cos(2\pi z)\,.
$$
The level sets $\fc_c$ have genus 3 and rank 3 for $c\in(-1,1)$ and genus 0
and rank 0 otherwise.
The function $\fc$ satisfies the property $T^*F=-F$, where $T$ is the translation
by $1/2$ in the three coordinate directions, so that when $\psi_a^*\fc=c$ has an
open level set then also $\psi_a^*\fc=-c$ has, meaning ultimately that
$\low{\fc}=-\upp{\fc}$ and so that, in particular, in order to study $\cD_\fc$
it is enough looking at the level $\fc_0$.
Dynnikov was able to find the analytical expression for the 10 largest
connected components of $\cD_\fc$ and their corresponding value of $\ell_\fc$.
\begin{figure}
  \centering
  \includegraphics[width=6.4cm]{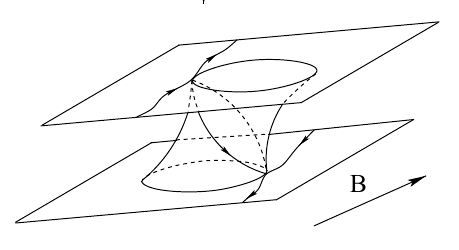}\hskip.4cm\includegraphics[width=6.4cm]{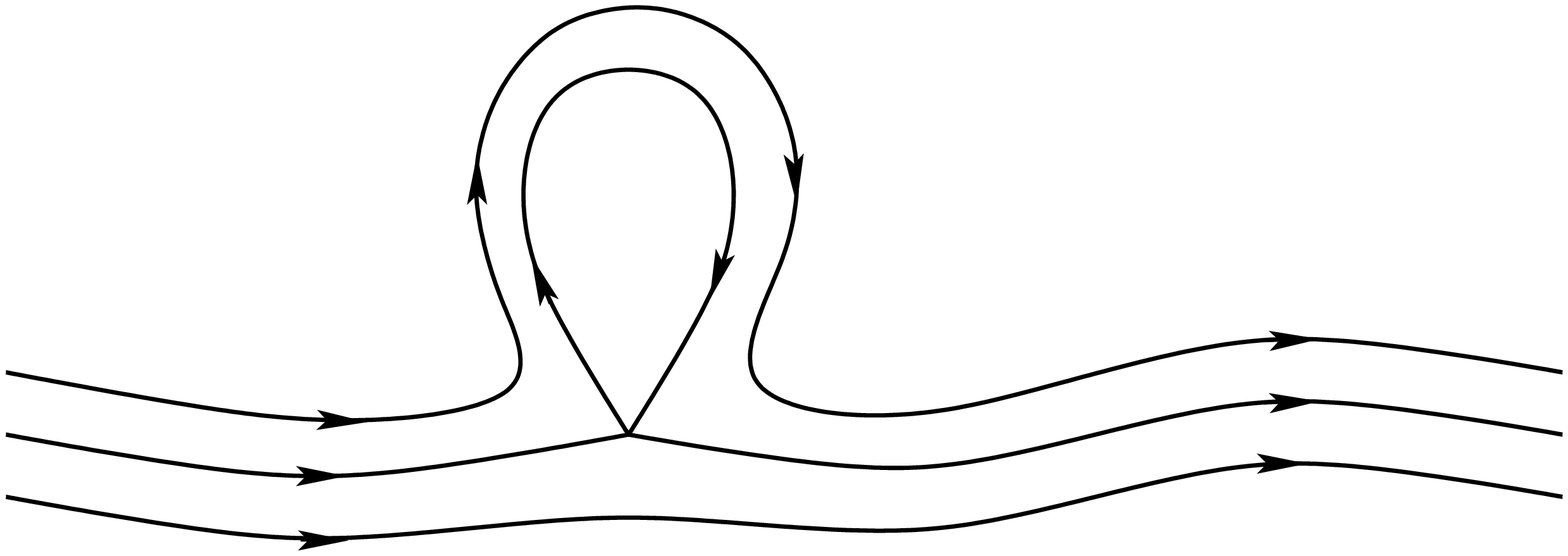}
  \caption{%
    \footnotesize
    (left) A typical critical section for a direction of the magnetic field at the boundary
    of a stability zone: the upper and lower bases of a cylinder of closed orbits
    collapse on each other creating a saddle connection. (right) A critical saddle
    point and behavior of trajectories near it.
  }
\label{fig:zero}
\end{figure}

The idea of this method is the following. Note first that, since $\c_0$ has
genus 3 and its curvature is strictly negative (except for eight points where
it is zero), all critical points, namely points where $\B$ is perpendicular
to the surface, are of saddle type (see Fig.~\ref{fig:zero}). Because of the Poincar\'e-Hopf index theorem,
the sum of the indices of a generic vector field on a manifold must equal its
Euler characteristics and so a generic (rational or irrational) direction
$\B\in\RPt$ must have exactly four of them: $4\cdot(-1)=2-2\cdot3$.
When $\B\in\QPt$ (that's the only case we can deal with numerically!),
generically each of these four saddle critical points is {\sl fully closed}, namely the
four tails are pairwise connected to each other forming two loops in such a way that
one of the loops is homotopic to zero while the second is not.

The four loops homotopic to zero are the bases of the two cylinders of closed
orbits -- recall that $\QPt\subset\cD_\fc$, so the open orbits induced by $\B$
are strongly asymptotic to a straight line and so the surface for each such
direction is decomposed into a pair of components $N_1,N_2$ and a pair
of cylinders $C_1,C_2$ (see~Fig.~\ref{fig:fol}).
Let $p_1\in\bR^3$ be one of the critical points. One can numerically find,
by trial and error, the coordinates of the critical point $p_2$ in $\bR^3$
at the opposite base of the cylinder. Clearly, in order for a new decomposition
of $\fc_0$ in cylinders and warped planes to arise it is necessary for the former cylinders
to collapse to zero height (see Fig.~\ref{fig:zero}), so {the equation $\langle p_1-p_2,\B\rangle=0$
defines a segment of the boundary of the stability zone where the original $\B$ lies.}
By following (numerically) the evolution of the zero-height cylinders,
at a certain point a second pair of critical points appears and the cylinders
will undergo a surgery. These points are vertices of the stability zones.
By following the lines and vertices defined by the new pairs until we get back
to the starting side, one ends up with the entire analytical boundary of a stability zone.
\begin{figure}
  \centering
  \includegraphics[width=5.4cm]{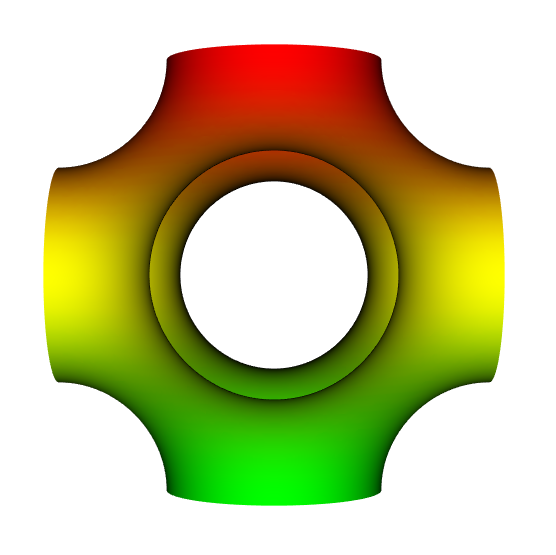}\hskip.76cm\includegraphics[width=5.4cm]{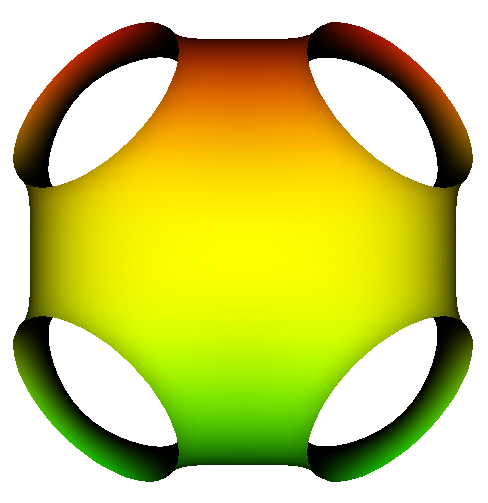}\\
  \vspace{.3cm}
  \includegraphics[width=5.4cm]{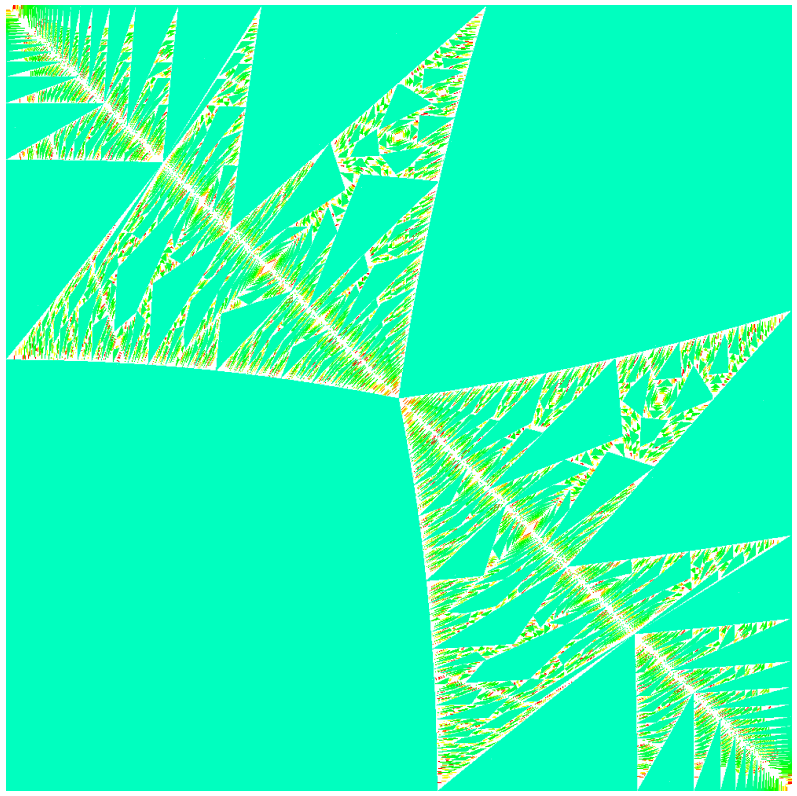}\hskip.76cm\includegraphics[width=5.4cm]{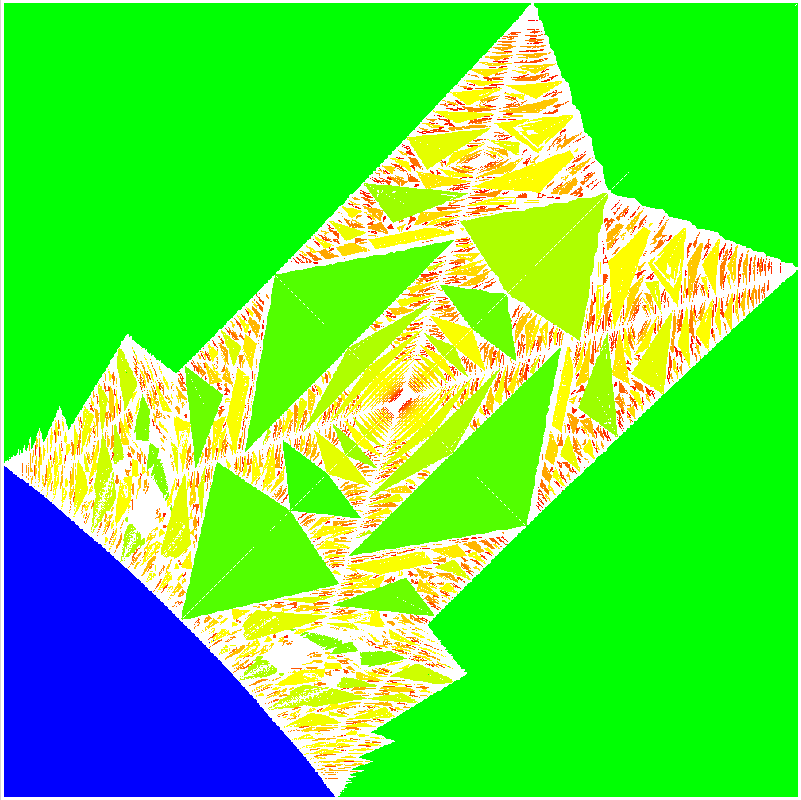}\\
  \vspace{.3cm}
  \includegraphics[width=5.4cm]{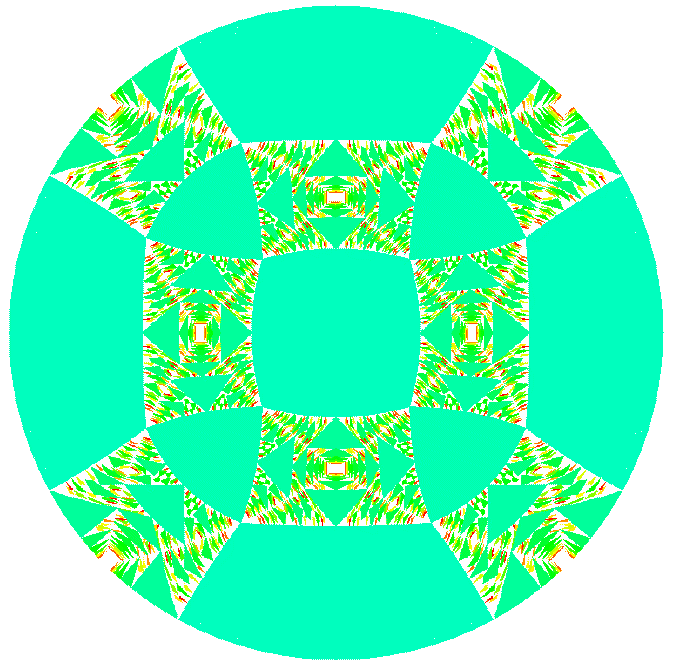}\hskip.76cm\includegraphics[width=5.4cm]{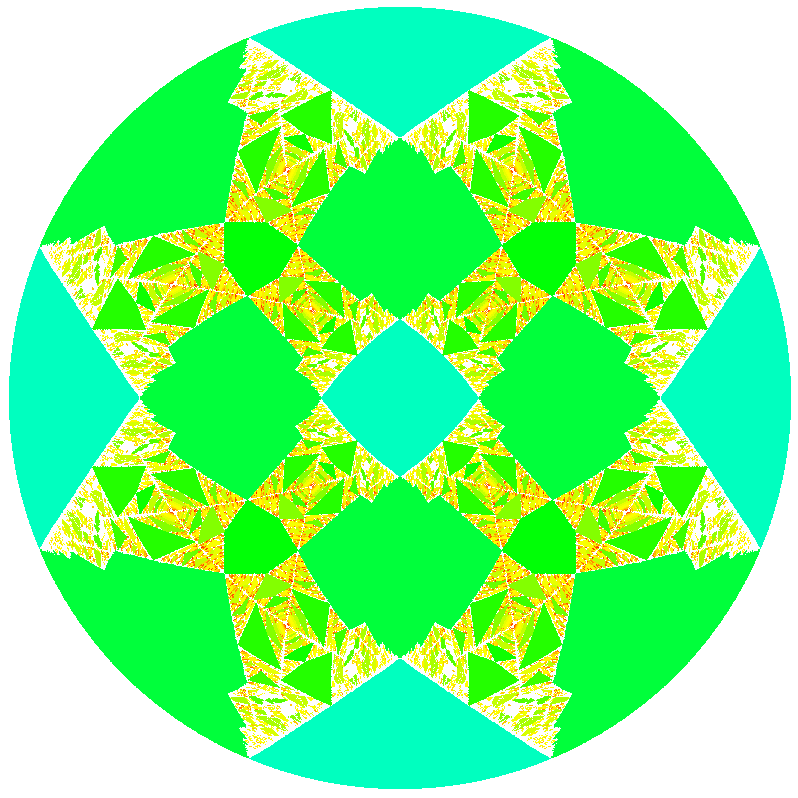}\\
  \caption{%
    \footnotesize
    (left) From top to bottom: the surface $\fc_0$, the SM $\cD(\fc)$ in the square $[0,1]^2$ of the chart $B_z=1$ and in the whole $\RPt$.
    (right) From top to bottom: the surface $\fd_0$, the SM $\cD(\fd)$ in the square $[0,1]^2$ of the chart $B_z=1$ and in the whole $\RPt$.
  }
\label{fig:cos3D}
\end{figure}

The drawback of this method is that it does not look suitable to be implemented
into a programming language. In order to bypass this problem, the first author
implemented the algorithm to evaluate $\ell_\fc(\B)$ in the open source C++
library NTC~\cite{NTC}. NTC is built on top of the open source C++ library
VTK by W. Schroeder, K. Martin and B. Lorense~\cite{SML06}, one of the most
popular computational geometry library
available online in the last two decades. VTK implements fundamental
geometry operations such as generating,  within some cuboid, the mesh for the
level set of a given function or generating the mesh of the intersection between
two such surfaces within some fixed cuboid.

While restricting an unbounded set to a bounded cuboid causes in general
a big loss of information, it is not so for a periodic set since the whole information
about it is contained inside a basic cell.
Surprisingly enough, in the authors' knowledge, none of the general-purpose
computational geometry libraries available online implement special algorithm
for {\sl periodic geometry}, although that is the only geometry where, quite
remarkably, ``{\sl it is possible to keep infinity inside a bounded box}''.
The NTC library implements exactly all periodic geometry algorithms described
above to evaluate $\ell_\fc(\B)$: finding the critical points induced by $\B$ on $\fc$;
retrieving the (whole) intersection (in $\bT^3$) between an embedded 2-torus
passing and a (periodic) surface; evaluating the homology of loops in $H_1(\fc,\bZ)$
and in $H_1(\bT^3,\bZ)$; finding the homology class in $\bT^3$ of a loop with a
given homology class in $\fc$. In order to get an approximation for $\cD_\fc$ and
$\ell_\fc$ with NTC it is enough to fix a grid in $\QPt$ and evaluate $\ell_\fc$
at all elements of the grid.

NTC currently supports functions with level surfaces of genus $g=3$ and $g=4$. The first is the simplest case with a
non-trivial set $\cD_\fc$, the second is the case of the Fermi surfaces of the noble metals Copper, Gold and Silver.
In Fig.~\ref{fig:cos3D} (left) we show $\fc_0$, the whole SM $\cD(\fc)$ and a detail of it in the region $[0,1]^2$
in the chart $B_z=1$.
A rough numerical evaluation of its box dimension gives an estimate of about 1.83, in agreement with Novikov's
Conjecture~1. 
In Fig.~\ref{fig:cos3D} (right) we show the set $\cD_\fd$ for the map 
$$
\fd(x,y,z)=\cos(2\pi  x)\cos(2\pi  y)+\cos(2\pi  y)\cos(2\pi  z)+\cos(2\pi  z)\cos(2\pi  x)\,,
$$
whose regular level sets $\fd_c$ are either spheres (for $c<-1$ and $c>0$) or genus-4 surfaces (for $-1<c<0$).
Each of the genus-4 level sets has topological rank 4. Note also that $\fd$, besides being invariant by integer translations
along the coordinate axes, is invariant with respect to translations by $1/2$ along the cube diagonals, namely it
has a bcc invariance. A rough numerical evaluation of its box dimension of about 1.69,
again in agreement with Novikov's Conjecture~1.
A striking confirmation of the correctness of these numerical data is shown in~\cite{DD09}. In that article
it is discussed the case of a simple piecewise linear function $F$ where the first author and Dynnikov were able
to find an analytical expression for $\ell_F$; the numerical data for that case agrees at 100\% level with the
analytical ones.

\begin{figure}
  \centering
  \includegraphics[width=4.25cm]{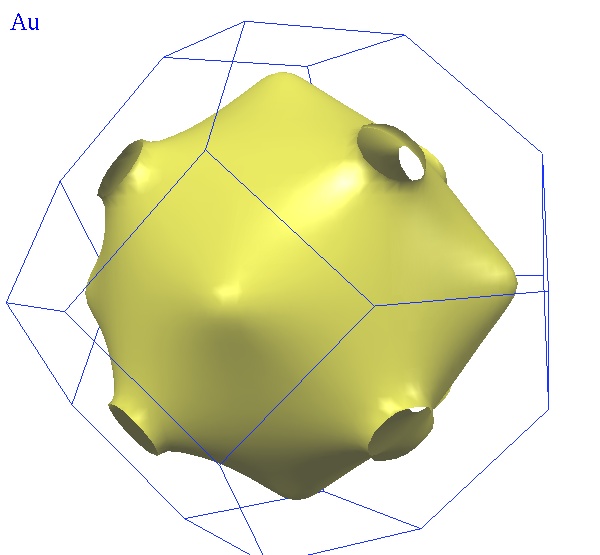}\hskip.40cm\includegraphics[width=4.25cm]{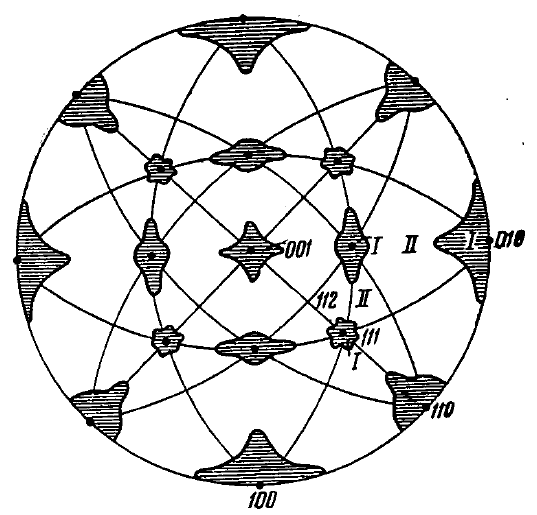}\hskip.40cm\includegraphics[width=4.25cm]{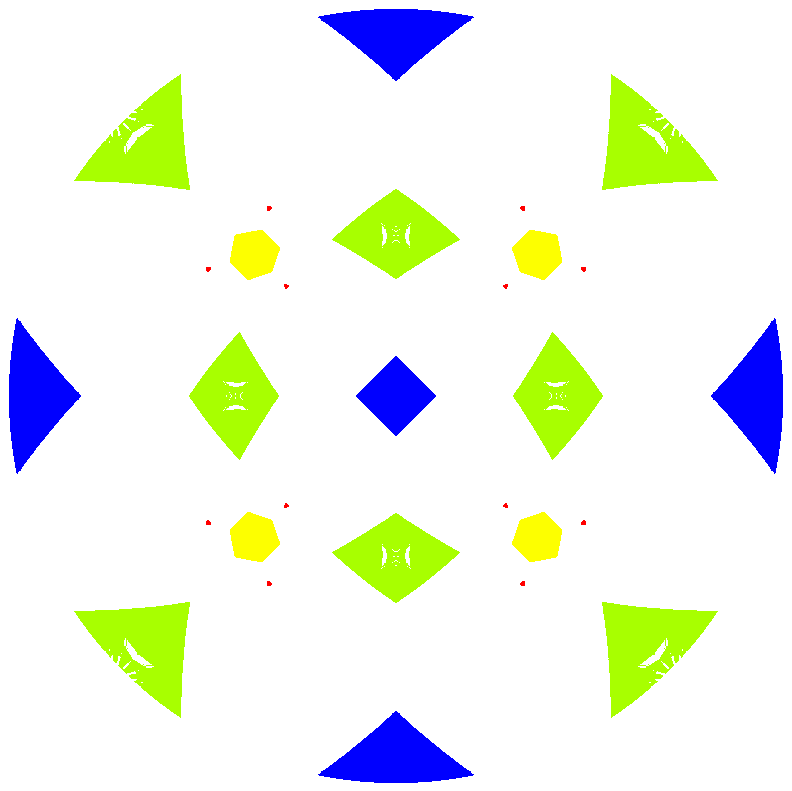}\\
  \vspace{.3cm}
  \includegraphics[width=4.25cm]{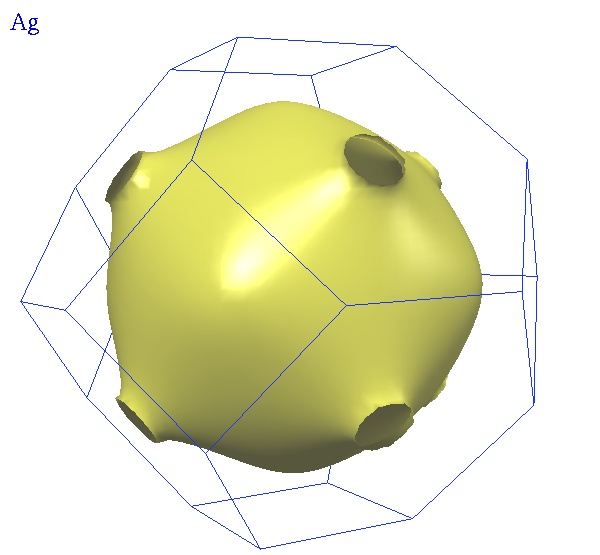}\hskip.40cm\includegraphics[width=4.25cm]{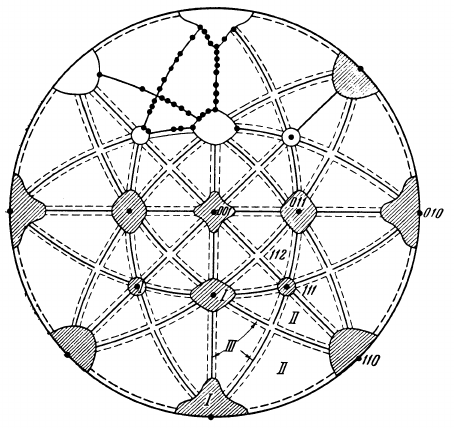}\hskip.40cm\includegraphics[width=4.25cm]{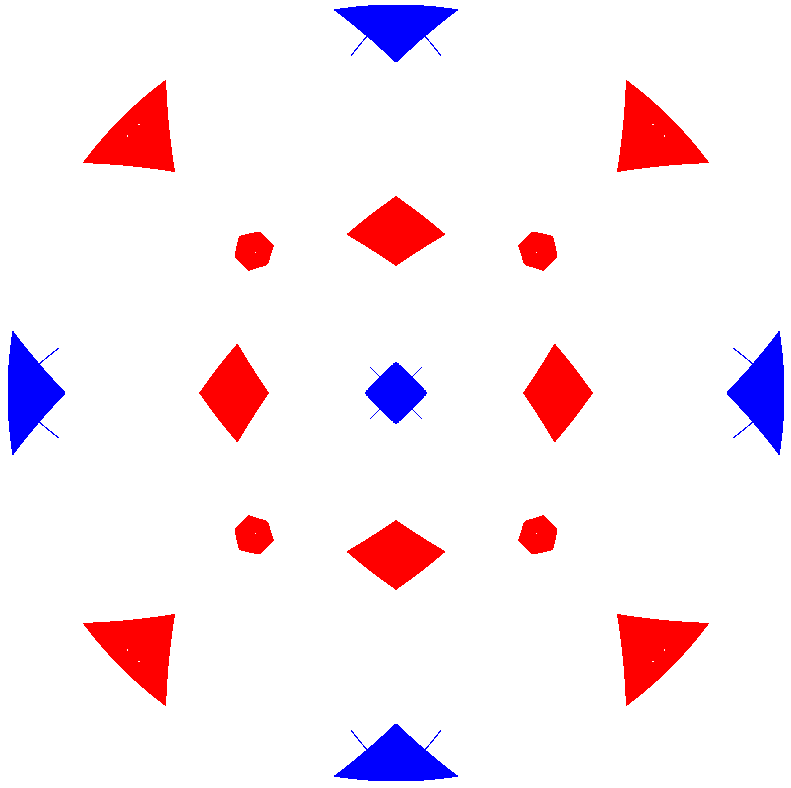}\\
  \vspace{.3cm}
  \includegraphics[width=4.25cm]{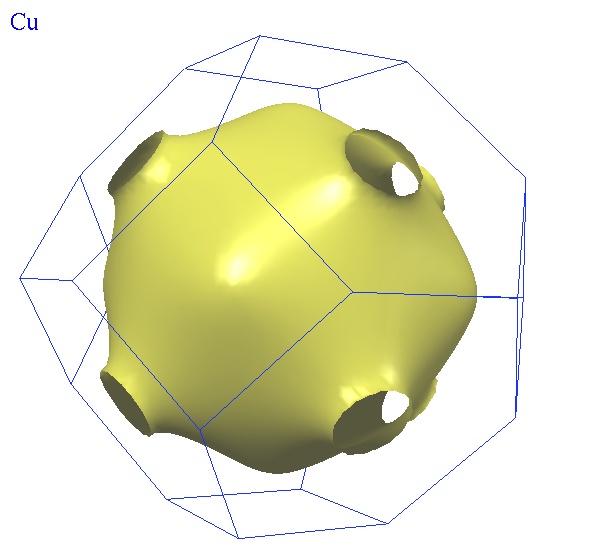}\hskip.40cm\includegraphics[width=4.25cm]{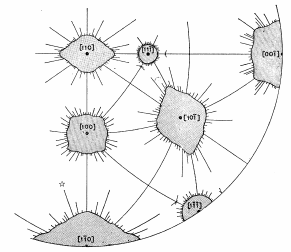}\hskip.40cm\includegraphics[width=4.25cm]{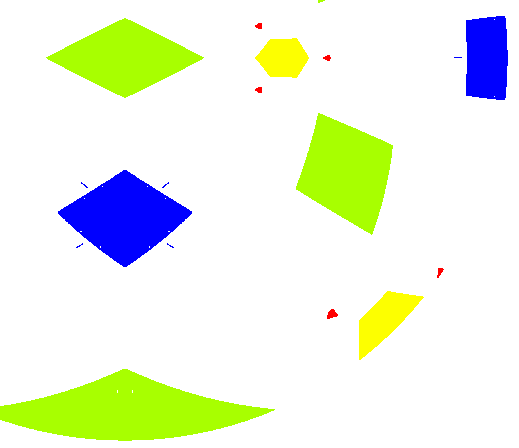}\\
  \caption{%
    \small
    (top) FS of Gold and its experimental~\cite{Ga60} and numerical~\cite{DeL04,DeL05b} Stereographic Map.
    (middle) FS of Silver and its experimental~\cite{AG62b} and numerical~\cite{DeL04,DeL05b} Stereographic Map.
    (bottom) FS of Copper and its experimental~\cite{KRBK66} and numerical Stereographic Map.
  }
\label{fig:nm}
\end{figure}

We switch now to the experimental data. As mentioned in the previous sections, according to the semiclassical approximation
the topology of the level sets of the QP function $\varepsilon_{\psi}$ given by the restriction of the Fermi energy function
to some plane $\psi$ perpendicular to $\B$ dictates the asymptotic behavior of the magnetoresistance for $\|\B\|\to\infty$
and so it can be detected experimentally. Starting from the end of Fifties, {\sl stereographic maps} were
experimentally obtained for many metals, mostly by Pippard, Alekseevskii and
Gaidukov~\cite{Pip57,AG59,Ga60,AG60,AGLP61,AG62a,AG62b,AG62c,AG63,AKM64}.
The maps for Gold, Silver and Copper are shown, respectively from top to bottom, in the middle column of Fig.~\ref{fig:nm}.
In these stereographic maps, $\RPt$ is represented as a disc and regions are shaded for those directions
of the magnetic field open orbits are detected and left blank otherwise. Mathematically,
this corresponds to the fact that we look only at a single level set $\varepsilon_{\psi}=c$, the Fermi energy
level, for every sibling of $\psi$ and, correspondingly, we define a {\sl reduced} map $\ell_{\varepsilon,c}(\B)$ that, for
any $\B\in\cD_{\varepsilon}$, is equal to $\ell_{\varepsilon}(\B)$ if $c\in[\low{\varepsilon},\upp{\varepsilon}]$
and to $(0,0,0)$ (meaning absence of open orbits) otherwise. We denote by $\cD_{\varepsilon,c}$ the subset of $\cD_{\varepsilon}$
where $\ell_{\varepsilon,c}(\B)\neq(0,0,0)$.

No comparison of these experimental data with theoretical prediction was possible for about half a century
because of the lack of knowledge about the levels of QP functions. In the right column of Fig.~\ref{fig:nm} we
show the numerical approximations of the sets $\cD_{\varepsilon,c}$ relative to approximated expressions
of the Fermi surfaces retrieved from the physics literature. Note that a strong magnetic field (of the order of 10 Tesla)
is needed in order for this phenomenon to become visible and these old experimental data was taken right
at the threshold (with magnetic fields of about 1 Tesla). Similarly, the trigonometric approximations we used
for the Fermi energy functions  is far from being the best approximation available to date (but it was the simplest
and quickest to implement in the NTC library). Yet, the match between experimental data and theoretical
prediction is remarkably high (see Fig.~\ref{fig:nm}).

We point out that the reason for using such old experimental data is that, after about a decade of great excitement that
saw a large number of theoretical and experimental articles dedicated to the subject, the interest of the solid state
community in the topic decreased a lot, possible exactly because no way was found to reproduce the experimental
data from first principles, and so in our knowledge no new stereographic maps were produced since the Sixties.
Recently, though, some new experimental result, in particular on the role of dislocations in the deformation of the
map $\ell_{\varepsilon,c}$ for Copper, has been published by M. Niewczas and his student Q. Bian~\cite{Bia10,BN16},
giving some hope for the appearance of accurate stereographic maps in a near future. Now that we have the
possibility, it would be indeed extremely interesting to have some more reliable experimental data to compare to. 

There are quite a few important advances to still make in this field:
\renewcommand\labelitemi{$\spadesuit$}
\noindent
\begin{itemize}
\item Extend NTC in order to make it work with surfaces of genus 7 (so is the Fermi surface of Lead~\cite{AG62a}).
\item Implement some algorithm in NTC able to produce accurate approximations of Fermi surfaces.
\item Explore sets $\cD_\varepsilon$ and $\cD_{\varepsilon,c}$ for several other triply periodic functions
  with level surfaces of genus 7 or less.
\item Optimize NTC to make it faster and more accurate.
\item Start the study the case of 4 and 5 quasiperiods.
\end{itemize}
\section*{Funding}
This material is based upon work supported by the National Science Foundation under Grant No. DMS-1832126
and the project ``Dynamics of complex systems'' (L.D. Landau Institute for Theoretical Physics).
\section*{Acknowledgments}
The authors are very grateful to S.P. Novikov for introducing the subject and for his constant
interest and support and also thank I.A. Dynnikov for many fruitful discussions on the subject over the years.
The numerical data in this article was produced on the HPCCs of the National Insitute of Nuclear Physics (INFN),
section of Cagliari (Italy), and of the College of Arts and Sciences (CoAS) of Howard University (Washington, DC).
\bibliography{refs}
\end{document}